\documentclass[11pt,showpacs,showkeys]{revtex4}
\input epsf.sty

\usepackage{ulem}
\usepackage{cancel}
\usepackage{color}

\def\comment#1{}

\usepackage{graphicx}
\begin{document}
\title{How universe evolves with cosmological and gravitational constants}
\author{She-Sheng Xue}
\email{xue@icra.it}
\affiliation{ICRANeT, Piazzale della Repubblica, 10-65122, Pescara,\\
Physics Department, University of Rome ``La Sapienza'', Rome,
Italy} 


\begin{abstract}
With a basic varying space-time cutoff $\tilde\ell$, 
we study a regularized and quantized Einstein-Cartan gravitational 
field theory and its domains of ultraviolet-unstable 
fixed point $g_{\rm ir}\gtrsim 0$ and ultraviolet-stable fixed point 
$g_{\rm uv}\approx 4/3$ of the gravitational gauge coupling 
$g=(4/3)G/G_{\rm Newton}$. Because the fundamental operators 
of quantum gravitational 
field theory are dimension-2 area operators,
the cosmological constant is inversely proportional to 
the squared correlation length
$\Lambda\propto \xi^{-2}$. 
The correlation length $\xi$ characterizes 
an infrared size of a causally correlate patch of the universe.
The cosmological constant $\Lambda$ and
the gravitational constant $G$ are related 
by a generalized Bianchi identity. 
As the basic space-time cutoff $\tilde\ell$ decreases and approaches 
to the Planck length $\ell_{\rm pl}$, the universe undergoes inflation 
in the domain of the ultraviolet-unstable fixed point $g_{\rm ir}$, 
then evolves to the low-redshift universe in 
the domain of ultraviolet-stable fixed point $g_{\rm uv}$.
We give the quantitative description of the low-redshift universe 
in the scaling-invariant domain of the ultraviolet-stable fixed point $g_{\rm uv}$, 
and its deviation from the $\Lambda$CDM can be 
examined by low-redshift $(z\lesssim 1)$ 
cosmological observations, such as supernova Type Ia. 
\end{abstract}



\maketitle

\section{\bf Introduction}

As one of fundamental theories for interactions in Nature, 
the classical Einstein theory of gravity,  which
plays an essential role in the standard model of modern cosmology ($\Lambda$CDM), 
should be realized in the scaling-invariant domain
of a fixed point of its quantum field theory, analogously to 
other renormalizable gauge field theories in the standard model of 
particle physics. It was suggested  \cite{w1} that the quantum 
field theory of gravity regularized at an ultraviolet (UV) cutoff 
might have a non-trivial 
UV-stable fixed point and asymptotic safety, namely 
the renormalization group (RG) flows are attracted 
into the UV-stable fixed point  
with a finite number of physically renormalizable operators 
for the gravitational field. 
The evidence of such UV-stable fixed point has been found 
in the different short-distance regularization frameworks of dimensional 
continuation \cite{w2}, the large $N$
(the number of matter fields) approximation \cite{w3}, 
lattice methods \cite{w4}, the truncated
exact renormalization group \cite{Reuter1998,w5,w7},
and perturbation theory \cite{w6}, as well as our approach to the quantum 
gravity \cite{ec_xue2010,xue2012g} that will be discussed below. 

The regularized and quantized gravitational 
field theory, see for example the bare action ${\mathcal A}_{\rm EC}$ of 
Eq.~(\ref{pact}) below, 
must be well-defined in a space-time and gauge independent 
regularization (cutoff) scheme for controlling quantum field fluctuating 
modes at short distances. 
In the scaling-invariant domain of the UV-stable fixed point, 
the operators of ${\mathcal A}_{\rm EC}$ obey RG-equations,
the irrelevant operators are suppressed by the cutoff, 
whereas the relevant operators become renormalizable 
operators at long distances. It is expected that the 
classical Einstein theory would be realized as an 
effective theory with two relevant and renormalizable operators
\begin{eqnarray}
{\mathcal A}^{^{\rm eff}}_{\rm EC}
&=&\int \frac{d^4x}{2\kappa}{\rm det}(e)(R-2\Lambda)+\cdot\cdot\cdot,
\quad \kappa\equiv 8\pi G,
\label{ec0}
\end{eqnarray}
where $R$ is the Ricci scalar, $\Lambda$ and $G$ are the cosmological 
and gravitational constant, and ``$\cdot\cdot\cdot$'' 
stands for matter fields and possible high-dimensional operators (see Ref.~\cite{kleinert}).  

It is a nature and important issue to study the asymptotic safety of
quantum gravitational field and their applications to 
the universe evolution. 
Some attempts have been made \cite{Reuter2002,weinberg2010}.  
We present our preliminary study of this issue in this article.

\section
{\bf The quantum gauge theory of gravitational field and its correlation length}\label{UV}

By the analogy of quantization of non-abelian gauge field theories,
we adopted \cite{ec_xue2010} the diffeomorphism and local (Lorentz)
gauge-invariant regularization scheme, which is a background-independent 
simplicial complex with a unique 
space-time running cutoff $\tilde\ell=\pi/\tilde\Lambda_{\rm cutoff}$, 
to regularize and quantize the Einstein-Cartan
field theory
for gravitational field, massless fermion and 
gauge fields \footnote{In this article, we do not explicitly show 
the matter section of fermion 
and gauge fields.}.
   
The basic gravitational variables in the Einstein-Cartan theory 
constitute a pair of tetrad and spin-connection fields 
$[e_\mu^{\,\,\,a}(x), \omega^{ab}_\mu(x)]$, whose Dirac-matrix values
are $e_\mu (x)= e_\mu^{\,\,\,a}(x)\gamma_a$
and $\omega_\mu(x) = \omega^{ab}_\mu(x)\sigma_{ab}$ with
Dirac matrices 
$\gamma_a$, $\sigma_{ab}$ and $\gamma_5$.
Analogously to the Wilson-loop defined in non-abelian gauge theories, 
we introduce the vertex field
\begin{equation}
v_{\mu\nu}(x)\equiv \gamma_5\sigma_{ab}(e^a_\mu e^b_\nu
-e^a_\nu e^b_\mu)/2,
\label{v}
\end{equation} 
and define the diffeomorphism and {\it local} Lorentz
gauge-invariant holonomy field \cite{ec_xue2010}
\begin{equation}
X_{\mathcal C}(e,\omega)=
{\mathcal P}_C{\rm tr}\exp\left\{ ig\oint_{\mathcal C}v_{\mu\nu}(x)
\omega^\mu(x) dx^\nu\right\},
\label{pa0s}
\end{equation} 
along the loop ${\mathcal C}$ on the four-dimensional Euclidean manifold, where 
$g$ is the gravitational gauge coupling.  
We introduced the 
$SO(4)$ group-valued spin-connection field
$U_\mu(x) = e^{ig\tilde\ell\,\omega_\mu(x)}$, and the smallest 
holonomy field $X_{h}(v,U)$ along the closed triangle path of
the 2-simplex $h(x)$ in the simplicial complex with the basic space-time cutoff  
$\tilde\ell$. We regularize and quantize 
the Euclidean Einstein-Cartan theory (see Eqs.~(120), (124) and (134) 
in Ref.~\cite{ec_xue2010}),
\begin{eqnarray}
{\mathcal A}_{\rm EC}(v,U)\!&=&\!\frac{1}{8g^2}\sum_{h\in {\mathcal M}}
X_{h} (v,U)
+{\rm H.c.},
\label{pact}
\end{eqnarray}
where $\sum_{h\in {\mathcal M}}$ is the sum over all 
2-simplices $h(x)$ of the simplicial complex, 
and the partition function 
\begin{eqnarray}
{\mathcal Z}_{\rm EC}[\tilde\ell,g(\tilde\ell)]
=\int{\mathcal D}e{\mathcal D}U\exp -{\mathcal A}_{\rm EC}. 
\label{part}
\end{eqnarray}
Henceforth, we will use ``the theory'' as the abbreviation of 
the regularized and quantized the Euclidean Einstein-Cartan 
field theory of Eqs.~(\ref{pa0s}), (\ref{pact}) 
and (\ref{part}). 
In this theory there are only two fundamental fields: 
the connection field-strength $g\,\omega_\mu$ and dimension-2
area-operator $\tilde\ell^2\,(e_\mu\wedge e_\nu)/2$ with two fundamental parameters:
the gravitational coupling $g$ and unique dimensional scale $\tilde\ell$. 
In the naive continuum limit of  $\tilde\ell\, g\,\omega_\mu \ll 1$ or 
$\tilde\ell\rightarrow 0$, 
Eq.~(\ref{pact}) formally reduces to Eq.~(\ref{ec0}) 
with the gravitational gauge coupling 
\begin{eqnarray}
g=(4/3)G/G_0, 
\label{g}
\end{eqnarray}
the cosmological 
constant $\Lambda=0$ and high-dimensional operators are naively 
suppressed by $\mathcal O[(\tilde\ell\, g\,\omega_\mu)^4]$. 
The $G$ is the running gravitational ``constant'' 
and $G_0$ is its present value, i.e., 
the Newton constant $G_{\rm Newton}=\ell^2_{\rm pl}$.

Instead of taking the naive continuum limit ($\tilde\ell\rightarrow 0$), 
one should first find the critical points of phase transitions. 
In the neighborhood of these critical points, 
one should then adopt the Kadanoff-Wilson approach to 
find an effective action at the scale of
the correlation length $\tilde\xi$ by integrating 
over short-distance modes at the space-time cutoff $\tilde\ell$.
Namely, in the simplicial complex of the cutoff $\tilde\ell$, one
integrates the partition function 
${\mathcal Z}_{\rm EC}[\tilde\ell,g(\tilde\ell)]$ 
of Eq.~(\ref{part}) over short-distance modes at the cutoff $\tilde\ell$ 
to obtain the effective partition 
${\mathcal Z}'_{\rm EC}[\tilde\ell\,',g(\tilde\ell\,')]$ 
in the simplicial complex of the cutoff $\tilde\ell\,'$, where $\xi> \ell'>\ell$. 
By comparing ${\mathcal Z}'_{\rm EC}[\tilde\ell\,',g(\tilde\ell\,')]$ 
to ${\mathcal Z}_{\rm EC}[\tilde\ell,g(\tilde\ell)]$, one could possibly 
obtain the running gravitational gauge
coupling $g(\tilde\ell)$, i.e., $\beta(g)$-function, and an 
effective quantum field theory 
in the continuum space and time. 

We integrate the partition function 
${\mathcal Z}_{\rm EC}[\tilde\ell,g(\tilde\ell)]$ 
of Eq.~(\ref{part}) over the short-distance quantum degrees of freedom $(e,U)$ 
of small loops of the length ${\mathcal C}_{\tilde\ell}\sim \tilde\ell$ and area 
$A({\mathcal C}_{\tilde\ell})\sim \tilde\ell^2$ so as to 
obtain an effective partition function 
${\mathcal Z}_{\rm EC}[\tilde\xi,g(\tilde\xi)]$ in terms of 
larger loops of the correlation length 
${\mathcal C}_{\tilde\xi}\sim \tilde\xi$ and 
area $A({\mathcal C}_{\tilde\xi})\sim \tilde\xi^2$ 
($\,\tilde \xi > \tilde\ell$). 
For the reason that the holonomy fields
$X_{h}(v,U)$ are dimension-2 area-operators in Eq.~(\ref{pact}),  
the leading contribution to the effective action 
${\mathcal A}_{\rm EC}[\tilde\xi,g(\tilde\xi)]$ contains 
a new volume term, i.e., the cosmological term 
\begin{eqnarray}
A({\mathcal C}_{\tilde\xi})/\tilde\xi^2\propto \sum_{{\mathcal C}_{\tilde\ell}} 
A({\mathcal C}_{\tilde\ell})/\tilde\xi^2 
\propto \sum_x \tilde\ell^4/(\tilde\ell^2\tilde\xi^2)
\propto \sum_x \tilde\ell^4\tilde \rho_{_{\tilde\Lambda}},
\label{vol}
\end{eqnarray}
where the energy density reads
\begin{eqnarray}
\tilde \rho_{_{\tilde\Lambda}}\propto 1/(\tilde\ell^2\tilde\xi^2),
\label{cmd}
\end{eqnarray}
and the related cosmological constant is given by
\begin{eqnarray}
\tilde\Lambda \propto \tilde\xi^{-2}.
\label{cmc}
\end{eqnarray}
Note that this preliminary discussion does not 
demand $\tilde\xi\gg \tilde \ell$ and
Eq.~(\ref{cmc}) can be $\tilde\ell$-dependent.

We have to confess that the result (\ref{cmc}) bases on the 
dimensional analysis. The rigorous calculations by both analytical and numerical approaches 
are necessarily required to obtain the cosmological term (\ref{vol}). 
This is a difficult task of the theory.
Nevertheless we argue that because the fundamental operators in the theory 
${\mathcal A}_{\rm EC}$ of Eq.~(\ref{pact}) are 
the dimension-2 area operators, rather than dimension-4 density operators, 
the dependence of the cosmological constant on 
the correlation length has the form of Eq.~(\ref{cmc}).

\section
{\bf The UV-unstable fixed point and its domain}\label{irpoint}

In this section, we discuss 
the UV-unstable fixed point $g_{\rm ir}$ and its domain ($g_{\rm ir}$-domain) 
of the theory of Eqs.~(\ref{pact}) and (\ref{part}).
The regularized action (\ref{pact}) at the space-time cutoff $\tilde\ell$
is actually the ratio of activation energy per the   
fundamental area operator $X_h$ and squared gravitational gauge 
coupling $g^2$, the latter acts as if it is a ``temperature''. 

In the weak coupling limit $g\approx 0$,  
all quantum fields of gravity and matter are asymptotically 
decoupled. Moreover, 
quantum fields $\{e,U\}$ are frozen to 
the configurations of completely randomly fluctuating and 
uncorrelated fields at the cutoff scale
$\tilde\ell$ and the fundamental operators $X_h$ have 
activation energies at 
the cutoff scale $\tilde\ell^{-1}$. This ``random'' 
configuration is a maximal-entropy configuration. It corresponds 
to the $\langle X_h\rangle=0$, where the expectational value 
$\langle \cdot\cdot\cdot\rangle$ is evaluated 
with respect to the partition function 
(\ref{part}). The reasons are that the partition function (\ref{part}), i.e., 
the amplitude of configurations $\{e,U\}$,
is dominated by the ``random'' configuration $\langle X_h\rangle=0$, and
other configurations with $\langle X_h\rangle\not=0$ 
are exponentially suppressed by their weight $\exp-\langle X_h\rangle/g^2$ 
for $g\approx 0$. 
This implies \cite{xue2012g} 
that (i) there is a disorder phase $\langle X_h\rangle=0$ 
at $g\approx 0$, where the fundamental operators $X_h$ 
are suppressed by their activation energies $\sim \tilde\ell^{-1}$, and 
(ii) the correlation between two fundamental operators $X_h$ is limited in 
their neighborhood and the correlation length $\tilde\xi\gtrsim\tilde\ell$.

As the coupling $g$ deviates from  $g\approx 0$ 
and increases for $g\gtrsim 0$, the more and more fundamental area operators 
$X_h$ start to activate and correlate each other. As a result 
the long-ranged order $(e,U)$-configurations 
of $\langle X_h\rangle\gtrsim 0$ are excited by the probability 
$\sim \exp-\langle X_h\rangle/g^2$. This means that in a length scale 
larger than the cutoff $\tilde\ell$, more and more 
degrees of freedom of quantum fields $\{e,U\}$ are correlated and the 
activation energy (per $X_h$) decreases. This implies the occurrence of a 
phase transition at the critical coupling 
\begin{eqnarray}
g_{\rm ir}=(4/3)(G_{\rm ir}/G_0)\gtrsim 0, \quad G_{\rm ir}\gtrsim 0.
\label{gir}
\end{eqnarray}
On the other hand, as the coupling $g$ increases and moves 
away from the critical coupling $g_{\rm ir}$, the ratio 
($\tilde\xi/\tilde\ell$) of 
the correlation length $\tilde\xi$ and UV-cutoff $\tilde\ell$ 
increases, indicating that the critical coupling 
$g_{\rm ir}$ should be a UV-unstable fixed point. 
This feature can also 
be seen from the perturbation 
result \cite{w6} of positive one-loop $\beta$-function 
$\beta(g_{_N})$, where $g_{_N}$ 
is a small dimensionless Newton constant in the 
effective theory of gravitational field with the UV-cutoff.
\comment{
$\beta(g_{_N})=2g_{_N} + {\mathcal O}(g_{_N}^2)>0
\mu(\partial \lambda/\partial \mu)=-2\lambda + {\mathcal O}(g_N)<0$
In addition, as will be discussed in Secs.~\ref{scalingf} 
and \ref{inflation}, the coupling $g$ and RG flows run away 
from the UV-unstable fixed point $g_{\rm ir}$, consistently with 
the space-time entropy increasing in time.
}

However, we know neither the exact $g_{\rm ir}$-value 
($g_{\rm ir}\approx 0$),
nor the measure and properties of the $g_{\rm ir}$-domain. 
Instead of being a scaling-invariant domain,
the part of $g_{\rm ir}$-domain is expected to be a crossover domain 
where the correlation length $\tilde\xi$ is not much larger than the
UV-cutoff $\tilde\ell$ so that 
$\tilde\xi$ does not decouple from $\tilde\ell$ variation. Therefore,
in addition to operators in the effective Einstein action (\ref{ec0}),
there are unsuppressed high-dimension operators of the regularized 
action (\ref{pact}) following the RG flows in the crossover domain.  
In this article, we postulate that  
the measure ($g$-range) of the crossover in the $g_{\rm ir}$-domain 
is small, namely, the ratio $\tilde\xi/\tilde\ell$ rapidly increases
for a small $g$-increase, corresponding to the positive
$\beta$-function 
$\beta(g)\equiv -\tilde\ell\partial g(\tilde\ell)/\partial\tilde\ell\approx 0^+$.
This postulation implies that for some small $g$-values 
in the $g_{\rm ir}$-domain, we would be allowed to make 
some approximate calculations by an expansion in the powers of 
$(\tilde\ell/\tilde\xi)<1$. We will come back to this point in the last section.  
Moreover, at the critical coupling $g_{\rm ir}$, 
the phase transition possibly occurs 
from the Planck phase ($g\lesssim g_{\rm ir}$), 
where all fundamental interactions 
including gravity are unified, 
to the phase ($g\gtrsim g_{\rm ir}$) 
of grant unification theory of gauge interactions to matter-fields.

\comment{We also present a brief discussion that as
the running space-time cutoff $\tilde\ell$ decreases possibly 
from the GUT scale $\tilde\ell=\pi/\tilde\Lambda_{\rm cutoff}$ 
($\tilde\Lambda_{\rm cutoff}\sim 10^{15}\,$ GeV), 
the RG flows of quantum gravity run away from the   
UV-unstable fixed point $g_{\rm ir}\approx 0$, resulting in the inflation 
(see Secs~\ref{irpoint} and \ref{inflation}). In addition, 
it will be discussed below that as the basic  
space-time cutoff $\tilde\ell$ {\it decreases}, 
the RG flows run away from the 
UV-unstable fixed point $g_{\rm ir}\approx 0$ of the theory. 
We should mention that the $g_{\rm ir}\approx 0$ is actually an infrared-stable 
fixed point of the theory, as the basic space-time cutoff $\tilde\ell$
of the theory {\it increase}, the RG flows are 
attracted into this fixed point.}

\section
{\bf The UV-stable fixed point and its domain}\label{uvpoint}

We briefly recall the UV-stable fixed point $g_{\rm uv}$ and its 
scaling-invariant domain 
($g_{\rm uv}$-domain) discussed in Ref.~\cite{xue2012g}.  
As the running space-time cutoff $\tilde\ell$ decreases, 
the increasing coupling $g$ approaches to $g_{\rm uv}$ 
($g\nearrow g_{\rm uv}$). In the $g_{\rm uv}$-domain, 
the large-loop holonomy 
fields $\langle X_{\mathcal C} \rangle$ of Eq.~(\ref{loop-par1})
are not suppressed because 
the smallest loops $X_h$ undergo condensation by jointing together 
side by side to form surfaces whose boundaries appears as large loops 
$X_{\mathcal C}$. The correlation length $\xi$
characterizes the size 
of dominantly non vanishing holonomy fields
\begin{eqnarray}
\langle X_{\mathcal C}\rangle &=&\int {\mathcal D}e{\mathcal D}U \,X_{\mathcal C}(e,U)
\exp -{\mathcal A}_{\rm EC}(e,U)\label{loop-par1}\\
&\sim& \int {\mathcal D}A({\mathcal C}) \,\exp \,-\,A({\mathcal C})/\xi^2,\label{loop-par1'}
\end{eqnarray}
where $\int {\mathcal D}A({\mathcal C})$ is 
the functional measure of all possible
surface-area $A({\mathcal C})$ bound by the large loop ${\mathcal C}$. 
As a result, the system undergoes a second-order phase 
transition at a critical coupling $g_{\rm uv}$. 
As the cutoff $\tilde\ell$ decreases, 
the increasing coupling approaches to the UV-stable fixed point $g_{\rm uv}$ 
($g\nearrow g_{\rm uv}$). In its domain of scaling invariance, 
the surface-area of large loops $X_{\mathcal C}$ proliferates and
becomes macroscopically large with the scaling invariant 
correlation length 
$\xi\gg \tilde\ell$ and area $\xi^2\gg \tilde\ell^2$.   
We approximately obtained the scaling law \cite{xue2012g} 
\begin{eqnarray}
\xi^2 \propto \tilde\ell^2g^2_{\rm uv}/(g_{\rm uv}^2 -g^2),~~
\xi \propto \tilde\ell/(g_{\rm uv} -g)^{\nu/2},
\label{loop-par2}
\end{eqnarray}
the critical coupling 
$g_{\rm uv}=(4/3)G_{\rm uv}/G_0\sim \mathcal O(1)$ and critical exponent 
$\nu= 1$ of the UV-stable fixed point $g_{\rm uv}$. 
Assuming that the coupling $g$ approaches 
the fixed point $g_{\rm uv}$, as the cutoff $\tilde\ell$ approaches the 
Planck length ($\tilde\ell \rightarrow \ell_{\rm pl}$), 
we obtained $g_{\rm uv}\approx (4/3)$  or $G\approx G_0\,\,$ [see Eq.~(\ref{g})].
The area law (\ref{loop-par1'}) represents 
the leading contribution to 
the partition function (\ref{part}) after integrating 
over vacuum-vacuum quantum fluctuations of small loops ${\mathcal C}_{\tilde\ell}$ 
at short distances $\tilde\ell$. The volume term 
$A({\mathcal C})/\xi^2$ in Eq.~(\ref{loop-par1'}) has the 
same form of the cosmological term in Eq.~(\ref{vol}).

In this scaling-invariant domain of the UV-stable fixed point
\begin{eqnarray}
g_{\rm uv}=(4/3)(G_{\rm uv}/G_0)\approx 4/3, \quad G_{\rm uv} \approx G_0\,,
\label{guv}
\end{eqnarray} 
the Kadanoff-Wilson approach leads to the RG equations 
for physically relevant and renormalizable operators 
of the effective Einstein theory (\ref{ec0}) at long distances. 
Applying the effective Einstein theory (\ref{ec0}) to the present
universe, 
we have (i) the gravitational coupling $G\approx G_0$ 
and the correlation length $\xi$ being 
the order of the size of the present universe horizon,  
(ii) the operator of Ricci scalar $R\propto \xi^{-2}$ and (iii)
the cosmological term 
\begin{eqnarray}
\Lambda \propto \xi^{-2},
\label{cmc1}
\end{eqnarray}
and the corresponding energy density 
\begin{eqnarray}
\rho_{_\Lambda}\propto (\tilde\ell \xi)^{-2},
\label{cmd1}
\end{eqnarray}
instead of $\rho_{_\Lambda}\propto (\tilde\ell)^{-4}$
\footnote{
This is reminiscent of the ``vacuum-energy'' density  
$\rho^{\rm vac}_{_\Lambda}\approx 1/(\ell^2_{\rm pl} \xi^2)$ rather than 
$\rho^{\rm vac}_{_\Lambda}\approx 1/(\ell^4_{\rm pl})$ as a candidate for 
the cosmological term, proposed by V.~G.~Gurzadyan and S.-S.~Xue 
more than 14 years ago, 
IJMPA 18 (2003) 561-568, see astro-ph/0105245 and astro-ph/0510495}. 
 
In the $g_{\rm uv}$-domain, the irrelevant operators 
of the theory (\ref{pact}) are suppressed 
by the powers of $(\tilde\ell/\xi)$, 
and the UV-cutoff $\tilde\ell$ is ``removed'' 
because of $\tilde\ell\ll \xi$. In the continuous spacetime, 
as a result, the Euclidean 
quantum or classical Einstein-Cartan field theory (\ref{ec0}) 
with finite numbers of relevant operators 
(e.g.~$\Lambda$ and $R$ terms) is realized as an effectively 
renormalizable theory in the sense of asymptotic safety.  
Such an effective Euclidean field theory is expected to become
an effective $3+1$ field theory (\ref{ec0}) 
after the Wick rotation \cite{ec_xue2010,xue2012g}.
The scaling-invariant correlation length $\xi$, 
equivalently the lowest lying mass scale $m=\xi^{-1}$ 
of quantum or classical gravitational fields, 
characterizes the infrared size ($L\propto \xi$) 
of a causally correlate patch of the universe.  

Note that the $g_{\rm uv}$-domain 
we chose the running space-time 
cutoff $\tilde\ell$ to approach the Planck length $\ell_{\rm pl}$ 
($\tilde\ell\rightarrow \ell_{\rm pl}$), 
rather than take 
the limit $\tilde\ell\rightarrow 0$ to remove the cutoff 
for the following reasons.  
The first, in a renormalizable field theory in the scaling-invariant 
domain of a non-trivial fixed  point, 
the removal of the cutoff $\tilde\ell$ 
means that renormalized physical operators at the scale $\xi$ 
do not depend on quantum field 
fluctuations at the cutoff $\tilde\ell\ll\xi$, the irrelevant
operators are suppressed by the powers of $(\tilde\ell/\xi)$.  
The second, the physical correlation length $\xi$ 
and coupling $g$ have to be determined by 
comparing final results with observations and experiments 
carried out at the known scales.
\comment{
, for example, the shortest Planck length $\ell_{\rm pl}=G_0^{1/2}$ 
The third, this choice 
does not preclude the possibility of
yet to known physics with even shorter 
length scale than the Planck scale and even longer length scale 
than the present Hubble length.}

Since fixed points and their scaling-invariant domains are 
universal, i.e., independent of different regularization 
schemes at the UV-cutoff, we expect that the 
nontrivial UV-stable fixed point 
$g_{\rm uv}$ of gravitational gauge coupling should relate 
to the non-Gaussian fixed point 
$(g^*_{_N},\lambda^*)$ of $g_{_N}=\tilde k^2G(\tilde k),
\lambda=\Lambda(\tilde k)/\tilde k^2$
for energy $\tilde k\rightarrow\infty$ obtained 
in Refs.~\cite{Reuter1998,w5,w7,w6}. 
However, in this article the cosmological constant appears as 
$\tilde\Lambda\propto \tilde\xi^{-2}$ 
of Eq.~(\ref{cmc}) or $\Lambda\propto \xi^{-2}$ of Eq.~(\ref{cmc1}),
relating to the scaling invariant correlation length of the theory (\ref{pact}), 
rather than a primary dimensional parameter following the RG equation 
in theory space of operators.

\section
{\bf The correlation length and scaling factor of FRW universe}\label{scalingf}

The correlation lengths $\tilde\xi$ and $\xi$ are defined by 
the two-point correlation function
of the theory (\ref{pact}) with the UV-cutoff $\tilde\ell$. 
They can be either the spatial length defined by 
the equal-time correlation function of two spatial points, 
or the time elapsing defined by the correlation function in time. 
In the domains of fixed points $g_{\rm ir}$ and $g_{\rm uv}$, an 
effective action of the Euclidean field theory of gravitation 
is obtained by integrating over a time-constant slice up to 
the size $\tilde\xi$ or $\xi$, and integrating over 
an imaginary time variable $\tau$ is up to the causally-correlate 
time $\tilde\xi$ or $\xi$. 
\comment{
An effective ``temperature'' $T_{_{\Lambda}}$ was introduced \cite{GH1977}
to characterize the cosmological term (\ref{ec0}), 
$T_{_{\Lambda}}\propto \Lambda^{1/2}$. 
From Eqs.~(\ref{cmc}) or (\ref{cmc1}) we then obtain 
\begin{eqnarray}
\tilde T_{_{\tilde\Lambda}} \propto \tilde\xi^{-1}\quad {\rm or} \quad
T_{_{\Lambda}} \propto \xi^{-1}.
\label{lt}
\end{eqnarray}
This means that the integration of imaginary time variable 
$\tau$ is up to $\tilde\xi$ or $\xi$ in the effective 
Euclidean finite-temperature field theory. 
}
Therefore, after the Wick rotating from the effective Euclidean  
field theory to the effective $3+1$ field theory 
for describing the universe evolution, 
the correlation lengths $\tilde\xi$ and $\xi$ are not only 
proportional to the infrared size $L$, but also proportional 
to the time elapsing $t$ of the expanding universe. We generally denote 
$\tilde\xi=\tilde\xi(t)$ and $\xi=\xi(t)$ monotonically 
increasing their values in time. 
The UV-cutoff $\tilde\ell=\tilde\ell(t)$ acts as a Lagrangian variable 
monotonically decreasing its value in time.
In fact, as a unique and arbitrary scale of the theory (\ref{pact}), 
the running UV-cutoff $\tilde \ell$ plays the roles of 
not only regularizing high-energy modes at short distances, 
but also providing a fundamental length unit measuring the 
correlation length $\tilde\xi/\tilde\ell$ or $\xi/\tilde\ell$.
The $\tilde\xi$ or $\xi$ turns out to be the physical unit 
for dimensional operators. 
The dynamics of the theory determine the coupling $g=g(t)$ 
and $\beta$-function $\beta[g(t)]$ as functions of time.  

In the flat ($k=0$) spatial section of the Robertson-Walker geometry 
with rotation and translation symmetries,
\begin{eqnarray}
ds^2=L^2(t)d\hat{\bf x}^2=a^2(t)d{\bf x}^2,
\label{ds2}
\end{eqnarray}  
where the spatial size $L(t)$ is the dimensional scaling factor, 
describing the length scale of the causally correlated universe 
at each time-constant slice in the four-dimensional space time. 
The dimensionless scaling factor is then given by 
$a(t)=L(t)/\tilde\ell(t)$, 
since the UV-cutoff $\tilde\ell(t)$ is a unique and basic 
scale of the theory (\ref{pact}). 
As discussed the dimensional scaling factor 
$L(t)\propto\tilde\xi(t)$ or $L(t)\propto \xi(t)$, 
\comment{
we have
\begin{eqnarray}
L(t)\propto \tilde T^{-1}_{_{\tilde\Lambda}}(t)~~~{\rm or}
~~~L(t)\propto T^{-1}_{_{\Lambda}}(t).
\label{enc}
\end{eqnarray}
Then}
and the dimensionless scaling factor of RW symmetry is given by 
\begin{eqnarray}
\tilde a(t)&=&L(t)/\tilde\ell(t)
\propto \tilde\xi(t)/\tilde\ell(t),
\label{al1}\\
a(t)&=&L(t)/\tilde\ell(t)
\propto \xi(t) /\tilde\ell(t)\, ,
\label{al2}
\end{eqnarray}
which is monotonically increasing in time, describes and measures 
the stretching of spatial manifold 
of space time in the universe expansion. Note that in Eq.~(\ref{al1}) 
we consider the postulation that the ratio $\tilde\xi/\tilde\ell$ becomes
sufficiently large in the part of $g_{\rm ir}$-domain so that we approximately 
adopt the effective Einstein theory (\ref{ec0}) and RW-geometry (\ref{ds2}).
We shall come back to this point in the last section.     

These discussions do not preclude the possibility that the correlation length
$\xi(t)$ of the theory discussed in this article describes only 
a part of the entirely causally correlated universe. In this case, instead
of a unique scaling factor $a(t)$ for the homogeneous and isotropic RW symmetry, 
we will probably be led to an effective cosmological term, 
which is contributed from the universe
inhomogeneity described by different correlation lengths 
at different parts of the universe. 
 
\comment{ 
Up to now, the correlation length $\tilde\xi$ or $\xi$ is the spatial length defined by 
the equal-time correlation function of two spatial points 
in the regularized Euclidean quantum field theory (\ref{pact}) 
with the running UV-cutoff $\tilde\ell$. 
In the domain of
fixed point $g_{\rm ir}$ or $g_{\rm uv}$, 
the $\tilde\ell$-dependence is removed 
for $\tilde\xi\gg \tilde \ell$ or $\xi\gg \tilde \ell$, 
a renormalizable quantum field theory or 
a ``static'' classical field theory in the continuous space time 
is reproduced with a proper Wick-rotation. 
The scale $\tilde\xi$ or $\xi$ is constant in time and plays role 
of an infrared (IR) cutoff $L$ 
of an effective quantum or classical field theory.
On the other hand, as a unique and arbitrary scale, the running UV-cutoff 
$\tilde \ell$ plays the roles of not only regularizing high-energy 
modes at short distances, but also providing a fundamental length unit measuring the 
correlation length $\tilde\xi/\tilde\ell$ or $\xi/\tilde\ell$. 
In the domain of fixed point $g_{\rm ir}$ or $g_{\rm uv}$, 
this is the scaling law $\tilde\xi/\tilde\ell=\tilde\xi[g(\tilde\ell)]$
or $\xi/\tilde\ell=\xi[g(\tilde\ell)]$ 
governed by the $\beta$-function, as 
will be discussed in the following sections.
The $\tilde\xi$ or $\xi$ turns out to be the physical unit 
for relevant physical operators in a renormalizable quantum field 
theory or ``static'' classical field theory in the continuous space time.
}

Moreover, conventional $3+1$ local quantum field theories 
with the IR-cutoff $L$ and the UV-cutoff $\Lambda_{\rm cutoff}$, whose 
entropy and energy scales extensively $S \sim L^3\Lambda_{\rm cutoff}^3$ 
and $E \sim L^3\Lambda_{\rm cutoff}^4$,  
vastly overcount degrees of freedom for a very large IR-cutoff. 
The reason is that these theories
are described in terms of Lagrangian volume-density operators, 
they have extensivity of the entropy built in. Therefore, it is required to have a
constrain on the IR- and UV-cutoffs, given by the up bound
energy \cite{ckn1999}    
\begin{eqnarray}
E\sim L^3\Lambda_{\rm cutoff}^4 \lesssim L M_{\rm pl}^2,
\label{ce}
\end{eqnarray}  
where the maximum energy density $\rho_{\rm max}\approx \tilde\Lambda_{\rm cutoff}^4$ 
and the Planck mass $M_{\rm pl}\propto 1/\ell_{\rm pl}$. When 
Eq.~(\ref{ce}) is near saturation, the maximum entropy is
\begin{eqnarray}
S_{\rm max} \approx (\pi L^2 M_{\rm pl}^2)^{3/4}.
\label{cs}
\end{eqnarray}
As discussed in Secs.~\ref{irpoint} and \ref{uvpoint},
the infrared cutoff $\tilde\xi$ and energy density 
$\tilde\rho_{_{\tilde\Lambda}}$ (\ref{cmd}) [or $\xi$ and $\rho_{_\Lambda}$ (\ref{cmd1})] 
satisfy the constrain (\ref{ce})
\begin{eqnarray}
\tilde  E_{_{\tilde\Lambda}}\approx 
\tilde\xi^3\tilde\rho_{_{\tilde\Lambda}}&\propto& \tilde\xi\, 
\tilde\ell^{-2} \lesssim \tilde\xi \,\ell^{-2}_{\rm pl},\label{ce1}\\
E_{_\Lambda}\approx \xi^3\rho_{_{\Lambda}} &\propto& \xi\, 
\tilde\ell^{-2}\lesssim \xi\,\ell^{-2}_{\rm pl}.
\label{ce2}
\end{eqnarray} 
They reach the saturation, when $\tilde\ell\rightarrow \ell_{\rm pl}$, 
corresponding the maximal entropy.
The reason is that the regularized Euclidean 
quantum gravity (\ref{pact}) is described in terms of area-density operators, 
rather than volume-density operators. In accordance with the energy density 
$\tilde\rho_{_{\tilde\Lambda}}$ (\ref{cmd}) or $\rho_{_\Lambda}$ (\ref{cmd1}), 
the entropy density 
$\tilde s_{_{\tilde\Lambda}}\approx 
\tilde\rho_{_{\tilde\Lambda}}\tilde\ell\propto 1/(\tilde\xi^2\tilde\ell)$
or
$s_{_\Lambda}\approx \rho_{_\Lambda}\tilde\ell\propto 1/(\xi^2\tilde\ell)$. The
the up bound entropy is 
\begin{eqnarray}
\tilde S_{_{\tilde\Lambda}}&\simeq &  \tilde\xi^3\,\tilde s_{_{\tilde\Lambda}} 
\propto \tilde\xi\, \tilde\ell^{-1}
\lesssim \tilde\xi\,\ell^{-1}_{\rm pl},\label{cso1}\\
S_{_\Lambda} &\simeq &  \xi^3\, s_{_\Lambda} \propto \xi\, \tilde\ell^{-1}
\lesssim \xi\,\ell^{-1}_{\rm pl}
\label{cso2}
\end{eqnarray}
which increases as $\tilde\xi$ or $\xi$ increasing and $\tilde\ell$ decreasing. 
This entropy is much smaller than the extensive entropy $\xi^3\tilde\ell^{-3}$ 
built by Lagrangian volume-density operators. These discussions show that the 
energy and entropy densities representing the cosmological term 
in Eq.~(\ref{ec0}) are contributed from 
the ensemble of quantum fluctuating degrees of freedom of 
the space time, which is described by the 
simplicial complex whose fundamental element 
is a 2-simplex, i.e., a triangle area operator 
at short distances $\tilde\ell$. 

\comment{
This can be regarded as an effective temperature of what people call the dark
energy. In this scenario of regularized Euclidean 
quantum gravity discussed in Secs.~\ref{UV} and \ref{uvpoint}, 
The entropy volume-density $\propto \tilde\ell^{-3}(t)$ or surface-density 
$\propto \tilde\ell^{-2}(t)$. The total entropy is 
$\propto (\tilde\xi[t,\tilde\ell(t)]/\tilde\ell(t))^{3}$ 
or $\propto (\tilde\xi[t,\tilde\ell(t)]/\tilde\ell(t))^{2}$,
considering the 
discussions of the possible relation of IR and UV 
cutoffs based on the entropy constrain of local quantum field 
theories \cite{ckn1999}.
The total entropy is conserved.}

As $\tilde\xi(t)$ or $\xi(t)$ increases and $\tilde\ell(t)$ 
decreases in time, the space-time entropy (\ref{cso1}) 
or (\ref{cso2})
\begin{eqnarray}
\tilde S_{_{\tilde\Lambda}}(t)\propto\tilde\xi(t)\tilde\ell^{-1}(t)
\propto \tilde a(t)~~{\rm or}~~ 
S_{_\Lambda}(t)\propto \xi(t)\tilde\ell(t)^{-1}\propto a(t)\label{ent}
\end{eqnarray}  
increases in time. This shows that $\tilde\xi(t)$ 
or $\xi(t)$ increasing 
and $\tilde\ell(t)$ decreasing in time are 
in accordance with the entropy of expanding universe increasing in time. 
As will be discussed, 
the proliferation and increase of space-time entropy 
$\tilde S_{_{\tilde\Lambda}}(t)$ and $S_{_\Lambda}(t)$ drive 
the universe inflation and acceleration. 
Whereas, the space-time energy Eq.~(\ref{ce1}) or (\ref{ce2}) 
\begin{eqnarray}
\tilde  E_{_{\tilde\Lambda}}(t)\propto\tilde\xi(t)\tilde\ell^{-2}(t)
\propto \tilde a(t)\tilde\ell^{-1}(t)~~{\rm or}~~ 
E_{_\Lambda}(t)\propto \xi(t)\tilde\ell^{-2}(t)
\propto a(t)\tilde\ell^{-1}(t),\label{eng}
\end{eqnarray}   
increases in time, as $\tilde\xi(t)$ or $\xi(t)$ increases 
and $\tilde\ell(t)$ decreases in time.   
This indicates that the matter-field sector (including dark matter) and 
the space-time sector (space-time originated cosmological term) 
should interact each other, resulting in
the energy exchange between the matter-field sector and space-time sector \cite{wangbin}.    

We expect that these two sectors must interact each other via 
microscopic particle-antiparticle (including dark matter particles) 
creation and annihilation. It is expected that 
the interaction of two sectors should be strong in 
the earlier universe evolution, and weak 
in the latter universe evolution.
\comment{
The temperature $T_{_{\Lambda}}$ 
is the energy scale of the space-time sector, and   
the temperature $T_{_{M}}$ of the matter-field sector, which
is assumed in the form of perfect fluid and thermal distribution.  
two sectors could be in the configuration of energy 
equipartition $\tilde T_{_{\tilde\Lambda}}\approx T_{_{M}}$.
the temperatures $T_{_{\Lambda}}$ and $T_{_{M}}$ 
should be approximately independent.
}      
These issues will be studied in future.
In this article, we will only use the 
generalized Bianchi identity (\ref{geqi00}) 
below for total energy-momentum conservation within the 
framework of classical 
Einstein equation for the universe evolution. 

\comment{
Considering the 
discussions of the possible relation of IR and UV 
cutoffs based on the entropy constrain of local quantum 
field theories \cite{ckn1999}, we generally define the cosmic
scaling factor from Eq.~(\ref{ir})
\begin{eqnarray}
a= a(\xi,\tilde\ell, \ell_{\rm pl}) \propto 
(\xi/\ell_{\rm pl})(\ell_{\rm pl}/\tilde\ell)^{\frac{2\alpha_1}{\alpha_1+\alpha_2}}.
\label{sa}
\end{eqnarray}
where $\alpha_1=$ and $\alpha_2=$. This general form does not affect 
the final result, except 
the coefficients and index value that should be determined by obs. 
and we 
adopt the simplest case for easy calculations.
}

In the following sections, we shall discuss the possible scenario 
that due to the proliferation and increase of the space-time entropy, 
the universe evolves in time. 
At the initial time $t=t_0$ and UV-cutoff
$\tilde\ell(t_0)=\tilde\ell_0$, 
the evolution starts with the inflation in the domain of 
the UV-unstable fixed point $g_{\rm ir}$ 
for the early universe. 
The universe evolution continues with acceleration in the 
scaling-invariant domain 
of the UV-stable fixed point $g_{\rm uv}$ 
for the present and future universe, as
the basic space-time UV-cutoff $\tilde\ell(t)$ decreases and 
approaches the Planck length $\ell_{\rm pl}$. 
We mainly consider that the evolution of low-redshift universe 
in the scaling-invariant
domain of the UV-stable fixed point $g_{\rm uv}$.
Using the scaling 
law (\ref{loop-par2}), we solve  
the Einstein equation for the cosmic scaling factor 
$a=a[\xi, g(\tilde\ell)]$ by taking into account 
the generalized Bianchi identity for the relation between the cosmological 
and gravitational constants $\Lambda=\Lambda[\xi,g(\tilde\ell)]$. 
In the last section, we 
give a brief and preliminary discussion 
on the inflation in the domain of the UV-unstable 
fixed point $g_{\rm ir}$ for the early universe.
This scenario is 
different from that  
discussed in Refs.~\cite{Reuter2002}, where the universe evolution 
follows the RG flow going away from the non-Gaussian fixed 
point $(g^*,\lambda^*)$ of high energies $(\tilde k\rightarrow\infty)$ 
to the RG-branch of low energies $(\tilde k\rightarrow 0)$ 
for the present universe. 

\section
{\bf The domain of UV-stable fixed point 
for the low-redshift universe}\label{present}

In the previous section, we mentioned that the $\xi(t)$ increasing 
and $\tilde\ell(t)$ decreasing in time consistently describe  
the universe expansion with entropy increasing in time. 
We further assume that the low-redshift universe 
has already been in the scaling-invariant domain of UV-stable
fixed point $g_{\rm uv}$. Namely, $\tilde\ell(t) \searrow \ell_{\rm pl}$ 
and $g(t) \nearrow  g_{\rm uv}$, physical 
quantities $m(g,\tilde\ell)$ are scaling-invariant and satisfy 
the renormalization-group invariant equation, i.e., 
$\tilde\ell (d m/d \tilde\ell)=0$, 
\begin{eqnarray}
\tilde\ell\frac{\partial m}{\partial \tilde\ell}
-\beta(g)\frac{\partial m}{\partial g}=0, 
\label{rge}
\end{eqnarray}
where the $\beta$-function is
\begin{eqnarray}
\beta(g)\equiv -\tilde\ell\frac{\partial g(\tilde\ell)}{\partial \tilde\ell} 
=\mu\frac{\partial g(\mu)}{\partial \mu},
\label{beta}
\end{eqnarray}
and the UV-cutoff $\mu\equiv\pi/\tilde\ell=\tilde\Lambda_{\rm cutoff}$.
In the neighborhood of the UV-stable fixed point $g_{\rm uv}$, 
where $\xi \gg \tilde\ell > \ell_{\rm pl}$, 
the coupling $g(\tilde\ell)$ and $\beta$-function 
can be generally expanded as a series 
\begin{eqnarray}
g(\tilde\ell)&=& g_{\rm uv} 
+c_0(\tilde\ell/\xi)^{1/\nu}+ {\mathcal O}[(\tilde\ell/\xi)^{2/\nu}],
\label{gexp}\\
\beta(g) &=& \beta(g_{\rm uv}) 
+ \beta'(g_{\rm uv}) (g-g_{\rm uv}) + {\mathcal O}[(g-g_{\rm uv})^2]>0,
\label{betaexp}
\end{eqnarray}
where $\beta(g_{\rm uv})=0$, $\beta'(g_{\rm uv})=-1/\nu$, the 
coefficient $c_0>0$
and the critical exponent $\nu>0$.  In the neighborhood of 
the fixed point $g_{\rm uv}$,
the behavior of the $\beta$-function 
\begin{eqnarray}
\beta'(g_{\rm uv})(g-g_{\rm uv})>0,~~~~~~{\rm for}~~~~ g<g_{\rm uv}
\label{betaf}
\end{eqnarray}
indicates the fixed point $g_{\rm uv}$ is UV-stable, 
as $\tilde\ell(t) \searrow \ell_{\rm pl}$ 
and $g(t) \nearrow  g_{\rm uv}$. 

Selecting the scaling-invariant physical quantity to be the 
correlation length $m=\xi^{-1}$, 
as the solution to 
Eq.~(\ref{rge}), we obtain that the correlation length $\xi$
follows the scaling law [cf.~ Eq.~(\ref{loop-par2})] 
\comment{for $\alpha_{1,2}=1$]
\begin{eqnarray}
\xi &= & \tilde\ell\exp +\int_{g_{\rm uv}}^g\frac{dg'}{\beta(g')}
=\ell_{\rm pl}
\left(\frac{\tilde\ell}{\ell_{\rm pl}}\right)^{\frac{2\alpha_1}{\alpha_1+\alpha_2}}
\left(\frac{c_0}{g_{\rm uv}-g}\right)^{\frac{\nu}{\alpha_1+\alpha_2}}
,\quad g<g_{\rm uv}, \label{ir}
\end{eqnarray}
}
\begin{eqnarray}
\xi (t)&= & \tilde\ell(t)\exp +\int_{g_{\rm uv}}^{g(t)}\frac{dg'}{\beta(g')},
\quad {\rm for}\quad g(t)<g_{\rm uv},\nonumber\\
&=&\tilde\ell(t)\left[\frac{c_0}{g_{\rm uv}-g(t)}\right]^{\nu/2},
 \label{ir}
\end{eqnarray}
which represents 
an intrinsic scale of the theory in the scaling-invariant domain. 
The dimensionless scaling factor
\begin{eqnarray}
a(t)\propto \xi(t)/\tilde\ell(t)
=\left[\frac{c_0}{g_{\rm uv}-g(t)}\right]^{\nu/2},
\quad {\rm for}\quad g(t)<g_{\rm uv}
\label{ira}
\end{eqnarray}
for the low-redshift universe. Using the scaling factor $a\equiv a(t)$ and 
gravitational coupling $g\equiv g(t)$, 
we will omit the time variable henceforth. 


We first introduce the scaling factor and gravitational coupling values 
at the present time $t_0$: $a_0$ and $g_0=g(a_0)\approx 4/3$ [cf.~ Eq.~(\ref{g})] 
for $G=G_0$. The relation (\ref{g}) between gravitational 
couplings $g$ and $G$ gives $(g/g_0)=(G/G_0)$.  
We consider the following evolution of universe in the scaling-invariant domain: 
\begin{description}
\item (i) $g \lesssim g_0 \lesssim g_{\rm uv}$ and $a\lesssim a_0 < a_c$ in the past;
\item (ii) $g_0 \lesssim g \lesssim g_{\rm uv}$ and $a_0 < a < a_c$ in the future,
\end{description}
where $a_c$ and $g_{\rm uv}=g(a_c)$ are the scaling factor and gravitational 
coupling values at the future time when the fixed point $g_{\rm uv}$ 
is approached $g(a_c)=g_{\rm uv}$.   

For two different values 
$a$ and $a_0$ of the scaling factor, the scaling law (\ref{ir}) yields 
\begin{eqnarray}
a^{2} \propto \frac{ (c_0)^\nu}{(g_{\rm uv}-g)^\nu};\quad
a^{2}_0 \propto  \frac{ (c_0)^\nu}{(g_{\rm uv}-g_0)^\nu},
\label{ir1}
\end{eqnarray}
and the ratio of two equations leads to the running gravitational coupling 
as a function of the scaling factor
\begin{eqnarray}
\left(\frac{g}{g_0}\right) &=&\left(\frac{g_{\rm uv}}{g_0}\right) + \left(1-
\frac{g_{\rm uv}}{g_0}\right)\left(\frac{ a}{a_0}\right)^{-2/\nu},\label{gair}
\\&\approx& 1+\delta_{_G} \ln (a/a_0)\approx 
\left(\frac{ a}{a_0}\right)^{\delta_{_G}}, \label{gair0}
\end{eqnarray}
where in the second line, $a\lesssim a_0$, $\ln (a_0/a)\ll 1$, and 
the parameter $\delta_{_G}\equiv (g_{\rm uv}/g_0-1)2/\nu>0$ 
is assumed to be small enough. Here, we treat $g_{\rm uv}$ and $\nu/2$ 
as parameters, hence the parameter $\delta_{_G}$,
to be fixed by observations. They can be in principle obtained by
non-perturbative calculations of $\beta$-function, including both 
gravitational and matter fields.  
The critical index $\nu/2$ actually relates to the anomalous 
dimension of the gravitational coupling $g$.

Eq.~(\ref{gair0}) shows that the low-redshift universe in the past 
$a(t)\lesssim a_0$, 
was approaching the present universe, 
$a(t)\nearrow a_0$ and $g(t)\nearrow g_0$. 
Eq.~(\ref{gair}) shows that the present universe is evolving 
into the universe in the remote future $a(t)/a_0\gg 1$ 
and $g(t)/g_0 >1$. Eq.~(\ref{gair}) also shows 
when the coupling $g(t)$ will be approaching
the UV-fixed point, $g(t)\nearrow g_{\rm uv}$, the universe will
be approaching its ``infinite'' size $a_c/a_0\rightarrow \infty$ in the 
``infinity'' time $t_c/t_0\rightarrow \infty$. 

\section
{\bf Einstein equation and generalized Bianchi identity}\label{Ein}

In the following sections, we will use the ``scaling'' 
relation (\ref{ir1}, \ref{gair}, \ref{gair0}) of 
scaling factor $a(t)$ and gravitational coupling $g(t)$ together with
the classical Einstein equations and total energy-momentum conservation
to study the evolution of low-redshift universe. 
Based on the observational facts and symmetry principle, at long distances
the Einstein tensor ${\mathcal G}_{ab}$ and the classical 
Einstein equation coupling to the total energy-momentum tensor $T_{ab}$ of matter fields 
can be in general written as (see for example \cite{wein1972} page 153)
\begin{eqnarray}
{\mathcal G}_{ab} = -8\pi G T_{ab}; \quad
{\mathcal G}_{ab} =  R_{ab} -(1/2) g_{ab}R -\Lambda g_{ab}.
\label{e1}
\end{eqnarray} 
The cosmological term $\Lambda g_{ab}$ in LHS of Einstein equation shows its 
gravitational origin.  
The covariant differentiation of Eq.~(\ref{e1}) and the Bianchi identity 
\begin{eqnarray}
[{\mathcal G}_{a}^{\,\,b}]_{\,;\,b} = -8\pi [G T_{a}^{\,\,b}]_{\,;\,b}, \quad
[ R^{\,\,b}_a-(1/2)\delta^{\,\,b}_aR]_{\,;\,b}\equiv 0,
\label{de1}
\end{eqnarray}
lead us to the generalized Bianchi identity: 
the conservation law of the energy-momentum of 
matter-field sector and space-time sector (the cosmological term)  
\begin{equation}
(\Lambda)_{;b}\,g^{\,\,b}_a=8\pi (G)_{;b}T^{\,\,b}_a+8\pi G 
(T^{\,\,b}_a)_{;b},
\label{geqi00}
\end{equation} 
where the cosmological and gravitational ``constants'' are no longer 
constant, i.e., $(\Lambda)_{;b}=(\Lambda)_{,b}$ and $(G)_{;b}=(G)_{,b}\,$. 
Using the result of Eqs.~(\ref{cmd}) and (\ref{cmc}) and 
$\tilde a\propto \tilde\xi/\tilde\ell$ for the early universe, 
or using the result of Eqs.~(\ref{cmc1}) and (\ref{cmd1}) and 
$a\propto \xi/\tilde\ell$ for the low-redshift universe, 
we in principle completely determine the relation of cosmological constant 
and gravitational coupling 
$\tilde\Lambda=\tilde\Lambda[\tilde a,g(\tilde\ell)]$ or
$\Lambda=\Lambda[a,g(\tilde\ell)]$ 
by the generalized Bianchi identity (\ref{geqi00}).

Suppose that the cosmological term $\Lambda\, g^{ab}$ does not exchange any 
mass-energy with matter fields for two possibly 
approximate cases: (i) the 
energy-density $\rho_{_\Lambda}$ is too small to energetically 
create many pairs of
particles and antiparticles in the low-redshift universe; (ii) 
the energy density $\rho_{_M}$ of particles and antiparticles 
is high and approximately comparable with $\tilde\rho_{_\Lambda}$ 
in the early universe. In these two cases, the conservation 
law of Eq.~(\ref{geqi00}) reduces to 
the energy-momentum conservation of matter fields 
$(T^{\,\,b}_a)_{;b}=0$ and relation 
\begin{equation}
(\Lambda)_{,b}=8\pi (G)_{,b}\, T^b_{\,\,\,\,a}=8\pi (G)_{,b}\, T^{bc} g_{ca},
\label{geqi0}
\end{equation} 
which relates the variations of $\Lambda$ and $G$ 
in the presence of matter fields. We practically study these 
two cases in the rest of this article. 

\section
{\bf Equation of state}\label{eos}

Matter fields in 
RHS of Einstein equation (\ref{e1}) 
are usually described by a perfect fluid with the energy-momentum tensor
\begin{equation}
T^{ab}=p_{_M}g^{ab} +(p_{_M}+\rho_{_M})U^aU^b,
\label{memt}
\end{equation}  
and the equation of state 
\begin{equation}
p_{_M}=(\Gamma_{_M}-1)\rho_{_M}\equiv\omega_{_M}\rho_{_M}, 
\label{meos}
\end{equation}
where $p_{_M}$ and $\rho_{_M}$ are the pressure and 
energy-density of matter field fluid, whose four-velocity $U^\mu$ 
obeys the condition $g_{\mu\nu}U^\mu U^\nu=-1$.    
The thermal index $\Gamma_{_M}>1$, i.e., $\omega_{_M}>0$ is due to
the facts that (i) the entropy of matter fields conserves, i.e., 
the number of particles conserves
in the universe expansion; (ii) the decreasing internal energy $E_{_M}$ 
(excluding mass-energy) of matter fields, i.e., 
$\delta E_{_M} =\delta(\rho_{_M}V)\leq 0$, paying for the universe expanding 
its volume $\delta V >0$. 
As a consequence of the particle-number and energy-conservation laws 
$(n_{\rm particle}U^b)_{;b}=0$ and $U_a(T^{ab}_{_M})_{;b}=0$ along a flow 
line in the universe expansion, we have
\begin{equation}
(\rho_{_M}U^b)_{\,;b} + p_{_M}U^b_{\,\,\,;b}=0.
\label{cmeq}
\end{equation}
Recalling that $dV/dt=VU^b_{\,\,;b}$, where $V$ and $t$ are the comoving 
volume and time, we have along each flow line  
\begin{equation}
p_{_M}\delta V +\delta E_{_M}=0, ~~ \Rightarrow ~~ p_{_M} \geq 0. 
\label{meqs}
\end{equation}  
Eq.~(\ref{meos}) and $\rho_{_M}>0$ lead to $\Gamma_{_M}-1=\omega_{_M}\geq 0$. 
The $\omega_{_M}$ value varies 
from $\omega_{_M}=1/3$ for ultra-relativistic matter fields to 
$\omega_{_M}=0$ for extremely non-relativistic matter fields. 

In order to understand the observational effects possibly due to  
the cosmological term $\Lambda g^{ab}$ in 
LHS of Einstein equation (\ref{e1}), 
one moves it to the RHS of Eq.~(\ref{e1}) and substitutes
it by an exotic ``dark energy'' fields
\footnote{Some confusions between the cosmological term 
and vacuum-energy were pointed out, see 
Section II of Ref.~\cite{xuecos2009}.}. 
By analogy with a perfect fluid of matter fields, 
one in general proposes the energy-momentum tensor
of the exotic ``dark energy'' fields to be
\begin{equation}
T^{ab}_{_D}\equiv
p_{_D}g^{ab} +(p_{_D}+\rho_{_D})U^aU^b
\label{cemt}
\end{equation}  
and the equation of state $p_{_D}=\omega_{_D}\rho_{_D}$. Moreover, 
for the case of $G\equiv G_0$ being constant 
the conservation law $(T^{ab}_{_D})_{;b}=0$ is demanded, independently
of the conservation law $(T^{ab}_{_M})_{;b}=0$ of matter fields \cite{review}.
Here we purposely use the subscript ``$D$'' to indicate
that as a kind of exotic matter field, 
this ``dark energy'' (\ref{cemt})
does not necessarily relate to 
the cosmological term $\Lambda g_{ab}$ in 
LHS of Einstein equation (\ref{e1}).
 
In fact, the cosmological term $\Lambda g^{ab}$ in 
LHS of Einstein equation (\ref{e1}) clearly 
has its gravitational origin as discussed above.  
If we move the cosmological term $\Lambda g^{ab}$ from 
LHS to RHS of Einstein equation (\ref{e1}), and rewrite it in the form
\begin{equation}
\Lambda\, g^{ab} \equiv - 8\pi G T^{ab}_{_\Lambda}, \quad T^{ab}_{_\Lambda}\equiv
p_{_\Lambda}g^{ab} +(p_{_\Lambda}+\rho_{_\Lambda})U^aU^b\label{c}.
\end{equation}  
As a result, the equation of state relating the pressure $p_{_\Lambda}$ 
and energy density $\rho_{_\Lambda}=\Lambda/(8\pi G)$: 
$p_{_\Lambda}=\omega_{_\Lambda}\rho_{_\Lambda}$, 
$\omega_{_\Lambda}\equiv -1$.
The conservation law $(T^{ab}_{_\Lambda})_{;b}=0$ follows for both 
$\Lambda\equiv \Lambda_0$ and $G\equiv G_0$ being constants.
In the case of both $G$ and $\Lambda$ dynamically varying, 
the generalized conservation law Eq.~(\ref{geqi00}) or (\ref{geqi0}) 
is fulfilled with 
the equation of state $p_{_\Lambda}=\omega_{_\Lambda}\rho_{_\Lambda}$
and $\omega_{_\Lambda}\equiv -1$.

Actually, our results of cosmological term $\Lambda\, g^{ab}$ 
and its properties presented
in Secs.~\ref{UV} and \ref{scalingf} are in agreement with 
the equation of state $p_{_\Lambda}=\omega_{_\Lambda}\rho_{_\Lambda}$
and $\omega_{_\Lambda}\equiv -1$. 
The negative pressure $p_{_\Lambda}<0$ or $\omega_{_M} < 0$ is due to
the facts that in the universe expansion 
$\delta V >0$, 
(i) the space-time entropy (\ref{ent}) increases,  
(ii) the space-time energy $E_{_\Lambda}$ (\ref{eng}) increases, i.e., 
$\delta E_{_\Lambda}  > 0$.  
This shows that the entropy and 
energy of stretching space-time manifold with a decreasing fundamental 
UV-cutoff $\tilde\ell$ are completely different from the entropy and 
internal (kinetic) energy of particles moving in the manifold \footnote{      
A physical interpretation  
was given in the last section of Ref.~\cite{xuecos2009}.}.

In the universe expansion, the space-time manifold in the form 
of the simplicial complex is stretched as the correlation length 
$\xi$ increases.
On the other hand, the fundamental element of the simplicial complex, i.e.,
a 2-simplex (a triangle area) 
of size $\tilde\ell$, becomes smaller. 
As a result the total entropy $S_{_\Lambda}$ (\ref{ent}) 
increases. This is an entropic repulsive force resulting 
from the entire universe's statistical tendency 
to increase its entropy.  
Against the gravitational attractive force, this entropic 
force acts in the universe and leads to its inflation and acceleration.
This entropic force is totally different from  
a particular underlying microscopic force of particles.

\section
{\bf The universe evolution with varying $G$ and $\Lambda$}\label{univ}

In this section, for the case of varying gravitational 
and cosmological ``constants'', $G$ and $\Lambda$, we study 
the Einstein equation (\ref{e1}) with the Robertson-Walker metric (\ref{ds2}) 
for the evolution of low-redshift universe. We implement the relation of the scaling factor 
$a(t)\propto\xi/\tilde\ell(t)$ of Eq.~(\ref{al2}) and the scaling law 
of Eqs.~(\ref{ira}-\ref{gair0}) in the scaling-invariant domain 
of the UV-stable fixed point $g_{\rm uv}$.  

Using the energy-momentum tensor (\ref{memt}) 
with $U^0=1$, $U^i=0$, $T^{00}=\rho_{_M}$ and 
$T^{ii}=p_{_M}g^{ii}$, the time-time and space-space components of the 
Einstein equation (\ref{e1}) in the Robertson-Walker metric are given by,
\begin{eqnarray}
3\ddot a &=& -[4\pi G (\rho_{_M}+3p_{_M}) -\Lambda]a, \label{e01}\\
a\ddot a +2\dot a^2 +2k &=& [4\pi G (\rho_{_M}-p_{_M}) +\Lambda]a^2. \label{e02}
\end{eqnarray}
\comment{
which become the Friedmann equation 
\begin{eqnarray}
(\dot a)^2 + k &=& \frac{1}{3} (8\pi G\rho_{_M} + \Lambda) a^2,\nonumber\\
H^2 &=& \frac{1}{3} (8\pi G\rho_{_M} + \Lambda)-\frac{k}{a^2}.
\label{f2}
\end{eqnarray}
}
As usually, the Hubble rate $H=\dot a/a$, 
the ``time''-variable $x=a/a_0 =1/(1+z)$ and 
$d(\cdot\cdot\cdot)/dt=(Hx)d(\cdot\cdot\cdot)/dx$. 
The values $z_0=0$ and $a/a_0=1$ represent the present time of the universe. 
The values $z>0$ and $a/a_0 <1$ 
represent the time in the past; the values $-1<z<0$ and $ a/a_0>1$ represent
the time in the future. 
Indicating the 
values of cosmological variables at the present time by subscript 
or superscript ``$0$'', we rewrite 
Eqs.~(\ref{e01}) and (\ref{e02}) as   
\begin{eqnarray}
H^2
&=& H_0^2\left(\frac{G}{G_0}\right)
\Big(\Omega_{_M} + \Omega_{_\Lambda}+\Omega_{_k}\Big),
\label{e3}\\
x\frac{dH^2}{dx}\! +\! 2H^2 \!&=&\!H^2_0\left(\frac{G}{G_0}\right)
\Big[2\Omega_{_\Lambda}\!-\!(1\!+\!3\omega_{_M})\Omega_{_M}\Big],\label{e2}
\end{eqnarray}
and the deceleration parameter
\begin{eqnarray}
q\!&\equiv&\! -\frac{(\ddot a a)}{\dot a^2}
\! =\!\frac{1}{2}\frac{(G/G_0)}{(H^2/H^2_0)}
\Big[\Omega_{_M}(1\!+\!3\omega_{_M})\! -\!2\Omega_{_\Lambda}  \Big],
\label{ed0}
\end{eqnarray}
where the conventional definitions of the critical density $\rho^0_c$,
\begin{eqnarray}
\rho^0_c=3H^2_0/(8\pi G_0),~~~~\Omega_{_{M,\Lambda,k}}=\rho_{_{M,\Lambda,k}}/\rho_c^0,
\label{ocri}
\end{eqnarray}
and the curvature density $\rho_k=-k/(8\pi G a^2)$. If the gravitational 
and cosmological ``constants'' are set to equal to their 
values at the present time, $G=G_0$  
and $\Lambda=\Lambda_0$, the above equations become usual equations 
in the $\Lambda\,$CDM. Moreover, at the present time,  
\begin{eqnarray}
\Omega^0_{_M}+\Omega^0_{_\Lambda}+\Omega^0_{_k}=1, \quad 
\Omega^0_{_{M,\Lambda,k}}=\rho^0_{_{M,\Lambda,k}}/\rho_c^0,
\label{omnow}
\end{eqnarray} 
and $\Omega^0_{_k}=-k/(H^2_0a_0^2)$ with the present values of energy densities 
$\rho^0_{_M}$, $\rho^0_{_\Lambda}= \Lambda_0/(8\pi G_0)$,
$\rho^0_k=-k/(8\pi G_0 a^2_0)$.  

In addition, the generalized Bianchi identity (\ref{geqi00}) becomes, 
\begin{eqnarray}
\frac{d\Lambda}{dt}
&=&-8\pi\left[\rho_{_M}\frac{dG}{dt}
+G\frac{d\rho_{_M}}{dt}+G\frac{3\dot a}{a}(p_{_M}+\rho_{_M})\right],
\label{cgeqi2}
\end{eqnarray}
or
\begin{eqnarray}
x\frac{d}{dx}\left(g\Omega_{_\Lambda}+g\Omega_{_M}\right)
&=&-3g(1+\omega_{_M})\Omega_{_M},
\label{cgeqi20}
\end{eqnarray}
where $g=G/G_0$, $g_0=1$ 
and Eq.~(\ref{omnow}) 
for $x_0=1$. In Eq.~(\ref{cgeqi20}), the coupling $g=g(x)$ is given
by the scaling law, for example Eq.~(\ref{gair}) if we know the 
$\beta$-function. However, we still need another independent equation relating 
$\Omega_{_\Lambda}$ and $\Omega_{_M}$ to find the solution 
$\Omega_{_\Lambda}(x)$ and $\Omega_{_M}(x)$ to Eq.~(\ref{cgeqi20}).
The relation between $\Omega_{_\Lambda}$ and $\Omega_{_M}$ should be
determined by the interaction of the space-time and matter-field sectors,
as already discussed after Eq.~(\ref{eng}) in Sec.~\ref{scalingf}. 
If this relation is known, we will be possibly able to 
find a resolution to the coincidence problem 
why $\Omega^0_{_\Lambda}$ and $\Omega^0_{_M}$ are
the same order of magnitude in the present universe. 

Analogously to the $\beta(g)$-function (\ref{beta}), 
using $(d/dt)=\dot{\tilde\ell}(d/d\tilde\ell)=\dot\mu (d/d\mu)$
and $\dot{\tilde\ell}\not=0$ $\dot\mu\not=0$,
we define the $\beta$-functions 
\begin{eqnarray}
\beta_{_\Lambda}
\equiv\mu \partial\Lambda(\mu)/\partial \mu,\quad 
\beta_{_M}\equiv\mu \partial\rho_{_M}/\partial \mu, \quad
\beta_a\equiv\mu \partial\ln a/\partial \mu.
\label{betas}
\end{eqnarray}
Eq.~(\ref{cgeqi2}) can be written as
\begin{eqnarray}
\beta_{_\Lambda}=-8\pi G_0[\rho_{_M}\beta(g) +g\beta_{_M} 
+3g\beta_a (p_{_M}+\rho_{_M})].
\label{bl1}
\end{eqnarray}
In the scaling-invariant domain of the UV-stable fixed point $g_{\rm uv}$,
as $\tilde \ell(t) \searrow \ell_{\rm pl}$ and $a(t)\nearrow a_0$, $\beta_a>0$, 
$\beta_{_M}< 0$ and $\beta(g) >0$, see Eq.~(\ref{beta}).

We turn to study the low-redshift universe, and suppose that the cosmological 
term and matter fields completely decouple from each other in this epoch. 
The energy-momentum conservation of matter fields
$(T^{\,\,b}_a)_{;b}=0$, i.e.,
\begin{eqnarray}
d(\rho_{_M} a^3)/da = -3 p_{_M} a^2, ~~ 
\Rightarrow~~\Omega_{_M}=\Omega^0_{_M}(a_0/a)^{3(1+\omega_{_M})}.
\label{mcl}
\end{eqnarray} 
The generalized Bianchi identity (\ref{geqi00})
reduces to (\ref{geqi0}). Eq.~(\ref{cgeqi2}) becomes 
\begin{eqnarray}
\frac{d\Lambda}{dt}
\!&=&\!-8\pi  \rho_{_M}\frac{dG}{dt}, ~~ {\rm or}~~
\frac{d(G\rho_{_\Lambda})}{dx}=-\rho_{_M}\frac{dG}{dx},
\label{geqi2}
\end{eqnarray}
and Eq.~(\ref{bl1}) becomes 
\begin{eqnarray}
\beta_{_\Lambda}=-8\pi G_0\rho_{_M}\beta(g)<0.
\label{bl2}
\end{eqnarray}
This indicates that the cosmological constant $\Lambda(t)$ decreases, i.e.,
$\Lambda(t) \searrow \Lambda_0$, 
as $a(t)\nearrow a_0$ and $\tilde \ell(t) \searrow \ell_{\rm pl}$
in the scaling-invariant domain of the UV-stable fixed point $g_{\rm uv}$.

Using Eqs.~(\ref{gair}) and (\ref{gair0}) in the scaling-invariant domain, 
with the boundary condition $x_0=1,\, a=a_0$ at the present time,  
we integrate the ``time''-variable $x=a/a_0$ in Eq.~(\ref{geqi2}) 
over the region 
$x\le 1$ for the past 
or the region $x\ge 1$ for the future,   
and obtain the evolution of cosmological constant 
\begin{eqnarray}
\frac{\Lambda}{\Lambda_0}
&=&\frac{G\Omega_{_\Lambda}}{G_0\Omega^0_{_\Lambda}}
=1-\left(\frac{\delta_{_\Lambda} }{\kappa}\right)
\left[1 - \left(\frac{a_0}{a}\right)^\kappa\right]
\label{lair}\\
&=&1-\delta_{_\Lambda}\ln x\sum_{n=0}^\infty\frac{(\kappa \ln x)^n}{(n+1)!}
\approx \left(\frac{a}{a_0}\right)^{-\delta_{_\Lambda}},
\label{lair0}
\end{eqnarray}
in the second line, $\ln (a_0/a)=\ln (1+z) \ll 1$ for the low-redshift universe.
The parameter $\kappa=3(1+\omega_{_M})-\delta_{_G}>0$.  The parameter 
$\delta_{_\Lambda}$ is related to the parameter $\delta_{_G}$ in the scaling 
law (\ref{gair0})
\begin{eqnarray}
\delta_{_\Lambda} =\delta_{_G} (\Omega^0_{_M}/\Omega^0_{_\Lambda})>0,
\quad \delta_{_\Lambda} <\delta_{_G}\ll 1
\label{deltagl}
\end{eqnarray}
due to the the generalized Bianchi identity (\ref{geqi2}). 
The small parameters
$\delta_{_G}$ and $\delta_{_\Lambda}$ should be determined 
and their relation (\ref{deltagl}) should be
checked by observational data of low-redshift universe.  
Eq.~(\ref{lair0}) shows that in the past $a\lesssim a_0$, 
$a\nearrow a_0$ 
and $\Lambda\searrow \Lambda_0$. 

For the case of low redshifts 
$z\lesssim \mathcal O(1)$, $\ln (a_0/a)=\ln(1+z)\ll 1$, and the 
indexes $\delta_{_G}<\delta_{_\Lambda}\ll 1$, Eqs.~(\ref{gair0}) 
and (\ref{lair0}) yield
\begin{eqnarray}
(G/G_0)
\approx(1+z)^{-\delta_{_G}},\quad (\Lambda/\Lambda_0)
\approx (1+z)^{\delta_{_\Lambda}},
\label{laira}
\end{eqnarray}
and $
(\Omega_{_\Lambda}/\Omega^0_{_\Lambda})
\approx (1+z)^{\delta_{_G}/\Omega^0_{_\Lambda}}$.
The correlation of gravitational and cosmological 
constants is
\begin{eqnarray}
(\Lambda/\Lambda_0)\!\approx\!(G/G_0)^{-(\Omega^0_{_M}/\Omega^0_{_\Lambda})};
~(\Omega_{_\Lambda}/\Omega^0_{_\Lambda})\!\approx\!
(G/G_0)^{-1/\Omega^0_{_\Lambda}},
\label{coir}
\end{eqnarray}
depending on the values $\Omega^0_{_\Lambda}$ and $\Omega^0_{_M}$. 
The ratio of $\Omega_{_M}$ and $\Omega_{_\Lambda}$ is 
\begin{eqnarray}
(\Omega_{_M}/\Omega_{_\Lambda})
\approx 
(\Omega^0_{_M}/\Omega^0_{_\Lambda})
(1+z)^{3(1+\omega_{_M})-\delta_{_G}/\Omega^0_{_\Lambda}},
\label{clm}
\end{eqnarray}
consistently with Eq.~(\ref{geqi2}). However, the interaction of 
space-time and matter-field sectors has been neglected.

Correspondingly, Eq.~(\ref{e3}) is approximately replaced by
\begin{eqnarray}
H^2
&\approx& H_0^2
\Big[\Omega_{_M}x^{+\delta_{_G}} + 
\Omega^0_{_\Lambda}x^{-\delta_{_\Lambda}}+\Omega^0_{_k}x^{-2}\Big],
\label{re3}
\end{eqnarray}
in contrast with the $\Lambda$CDM equations 
$(\delta_{_G} =\delta_{_\Lambda}=0$), the same discussions apply
for the deceleration parameter (\ref{ed0}) and the luminosity 
distance 
\begin{eqnarray}
d_H(z)=\int_0^{a_0/(1+z)}dz'/(1+z')H(z'),
\label{cdis}
\end{eqnarray}
where $H(z)$ is given by Eq.~(\ref{re3}).
The relations (\ref{deltagl}) and (\ref{re3}) 
can be examined \cite{xueyang2014} by 
using the measurements of low-redshift 
($z\lesssim 1$) cosmological observations, e.g.~Type Ia supernovae.

\begin{figure}
\begin{center}
\includegraphics[height=1.5in]{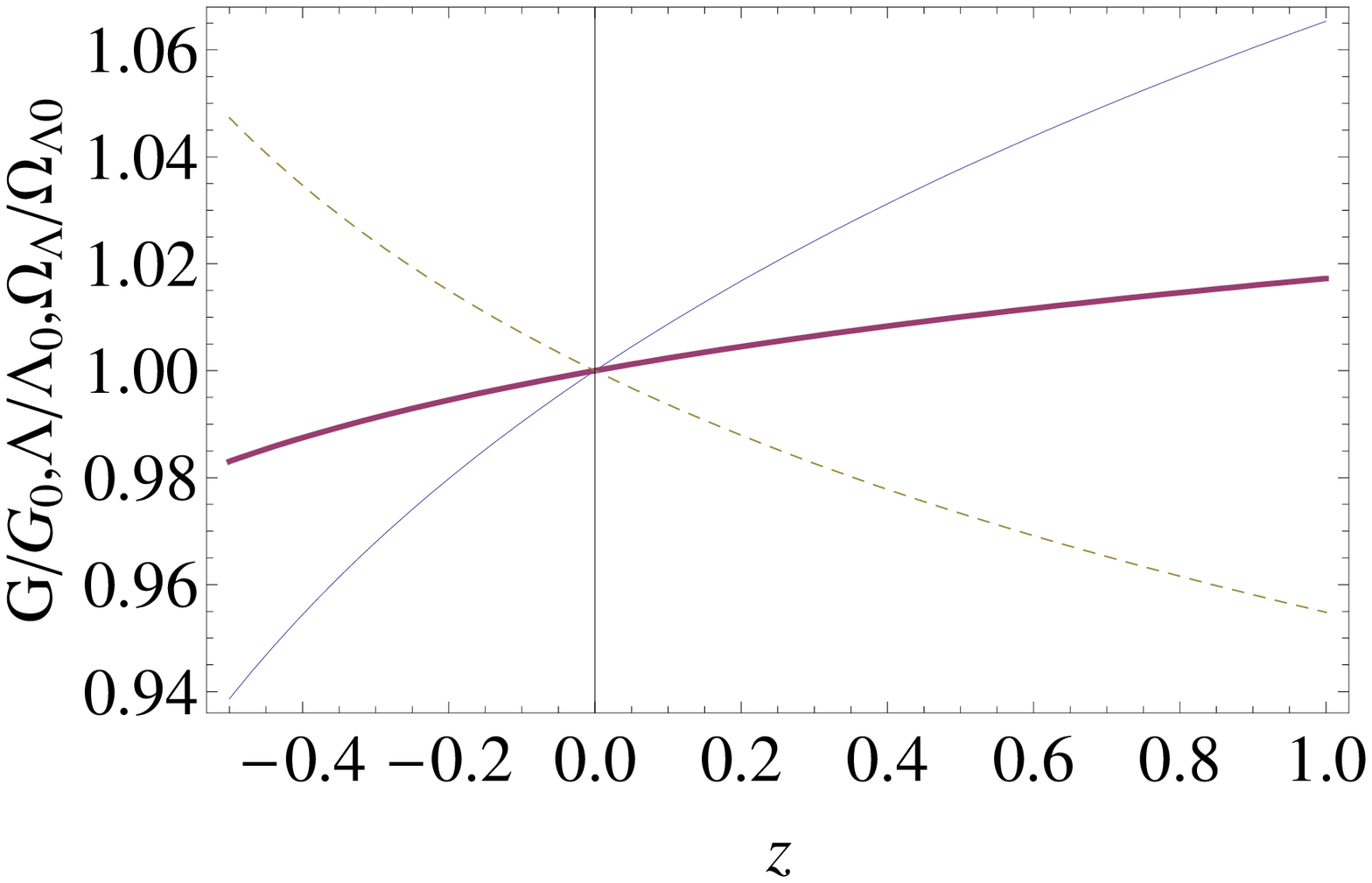}
\includegraphics[height=1.5in]{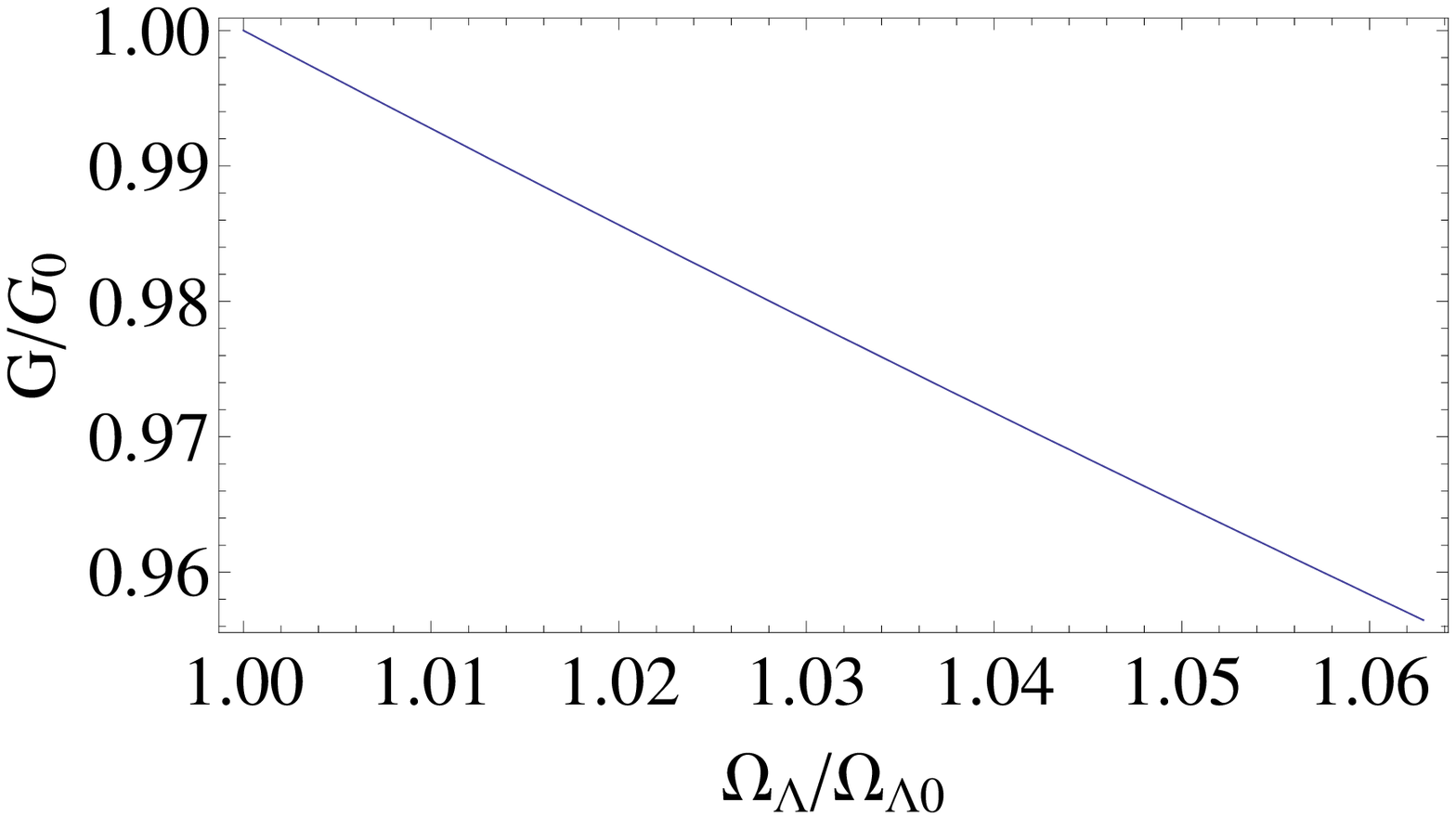}
\caption{As functions of the redshift $z$, we illustrate (i) 
$G/G_0$ (dashed line), $\Lambda/\Lambda_0$ (thick solid line) and  
$\Omega_{_\Lambda}/\Omega^0_{_\Lambda}$ 
(thin solid line) of Eq.~(\ref{laira}); (ii) the 
correlation (\ref{coir}) of the gravitational and cosmological constants.} \label{rgl}
\end{center}
\end{figure}

\begin{figure}
\begin{center}
\includegraphics[height=1.6in]{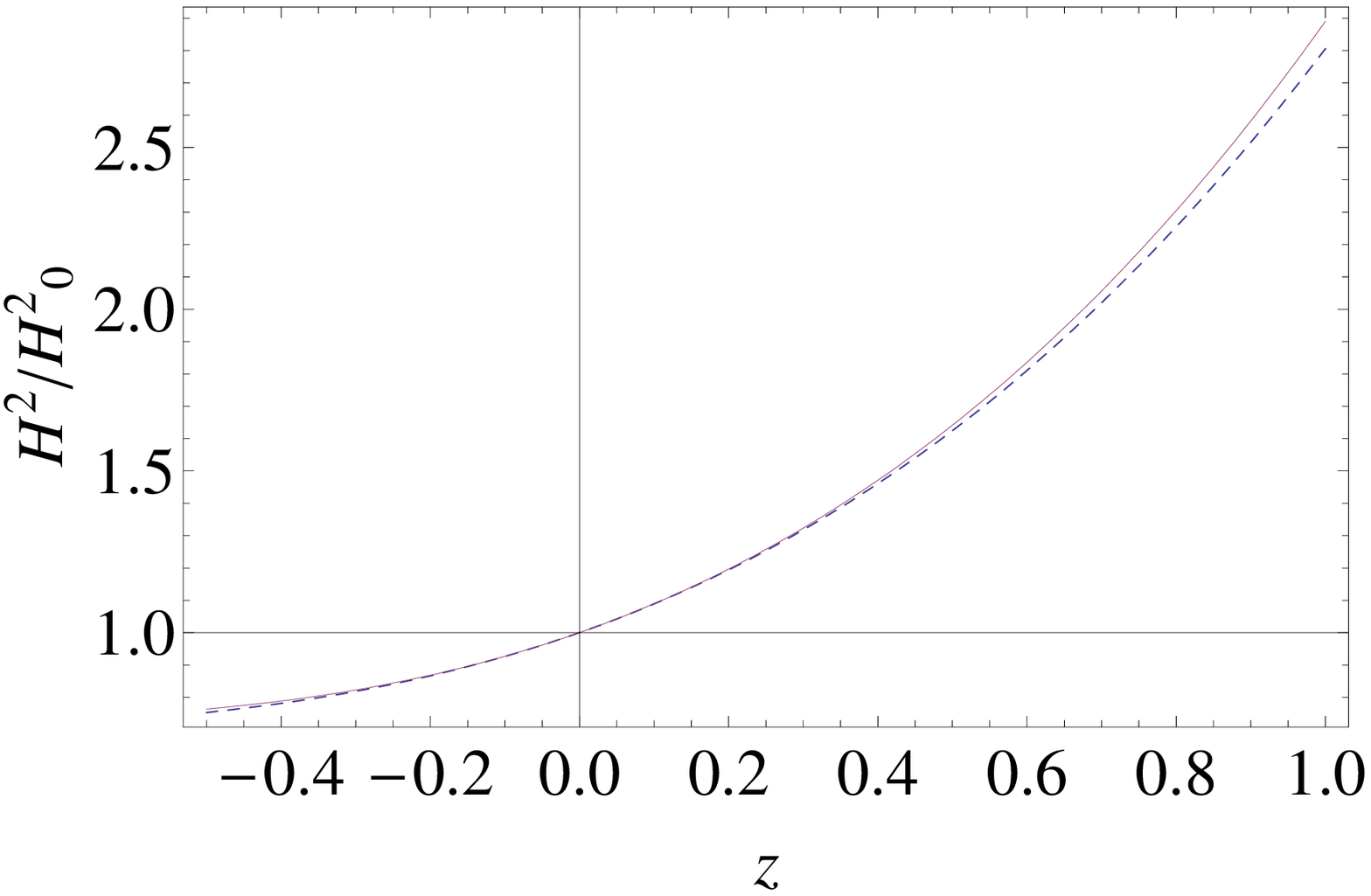}
\includegraphics[height=1.6in]{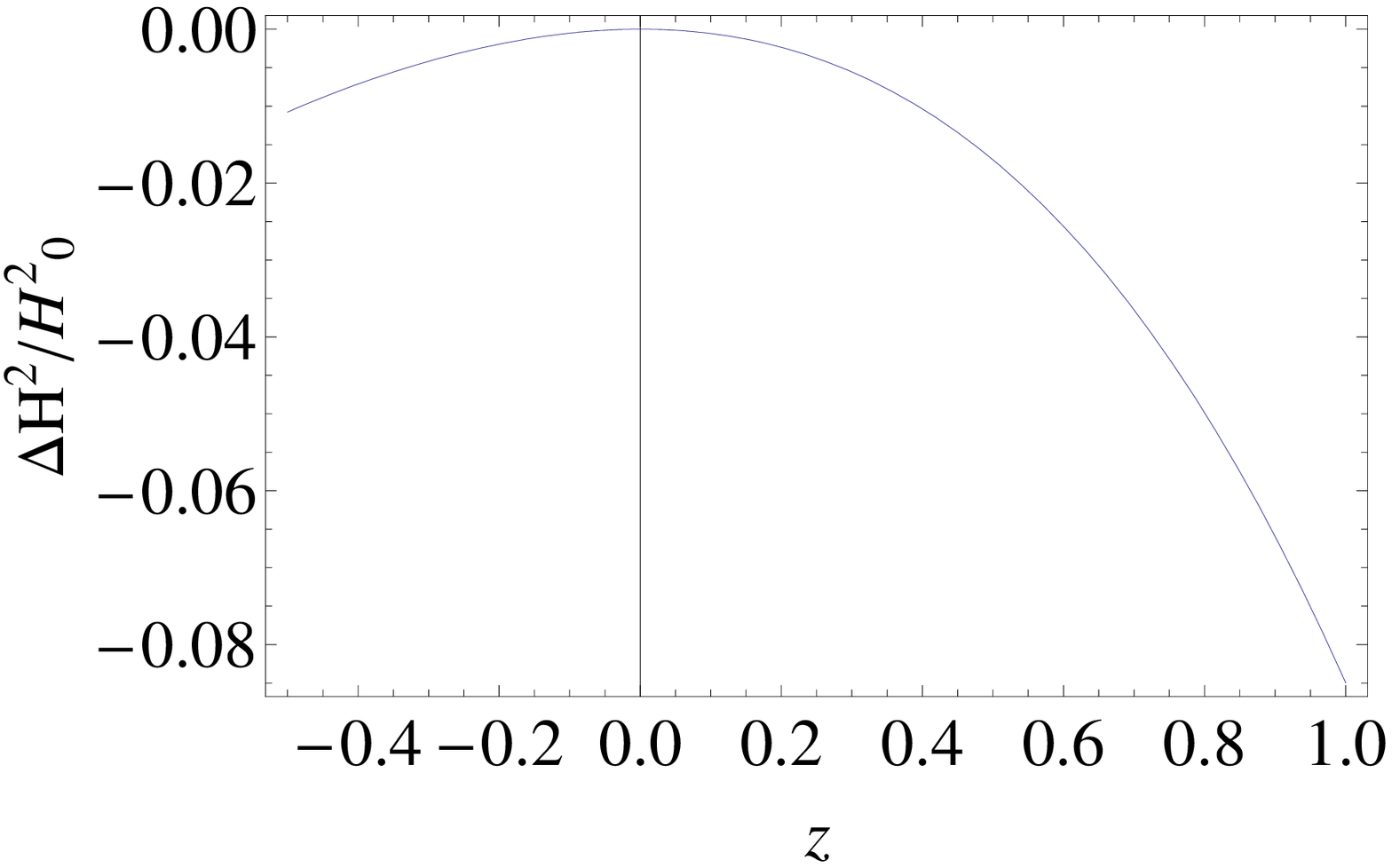}
\caption{We illustrate the 
Hubble function $H^2/H^2_0$ (dashed line) of Eq.~(\ref{re3}) in contrast 
with its $\Lambda$CDM counterpart (solid line), and their discrepancy 
$\Delta H^2\equiv H^2(\delta_{_G})-H^2(0)$ 
in terms of the redshift $z$.} \label{etimeh}
\end{center}
\end{figure}

\begin{figure}
\begin{center}
\includegraphics[height=1.5in]{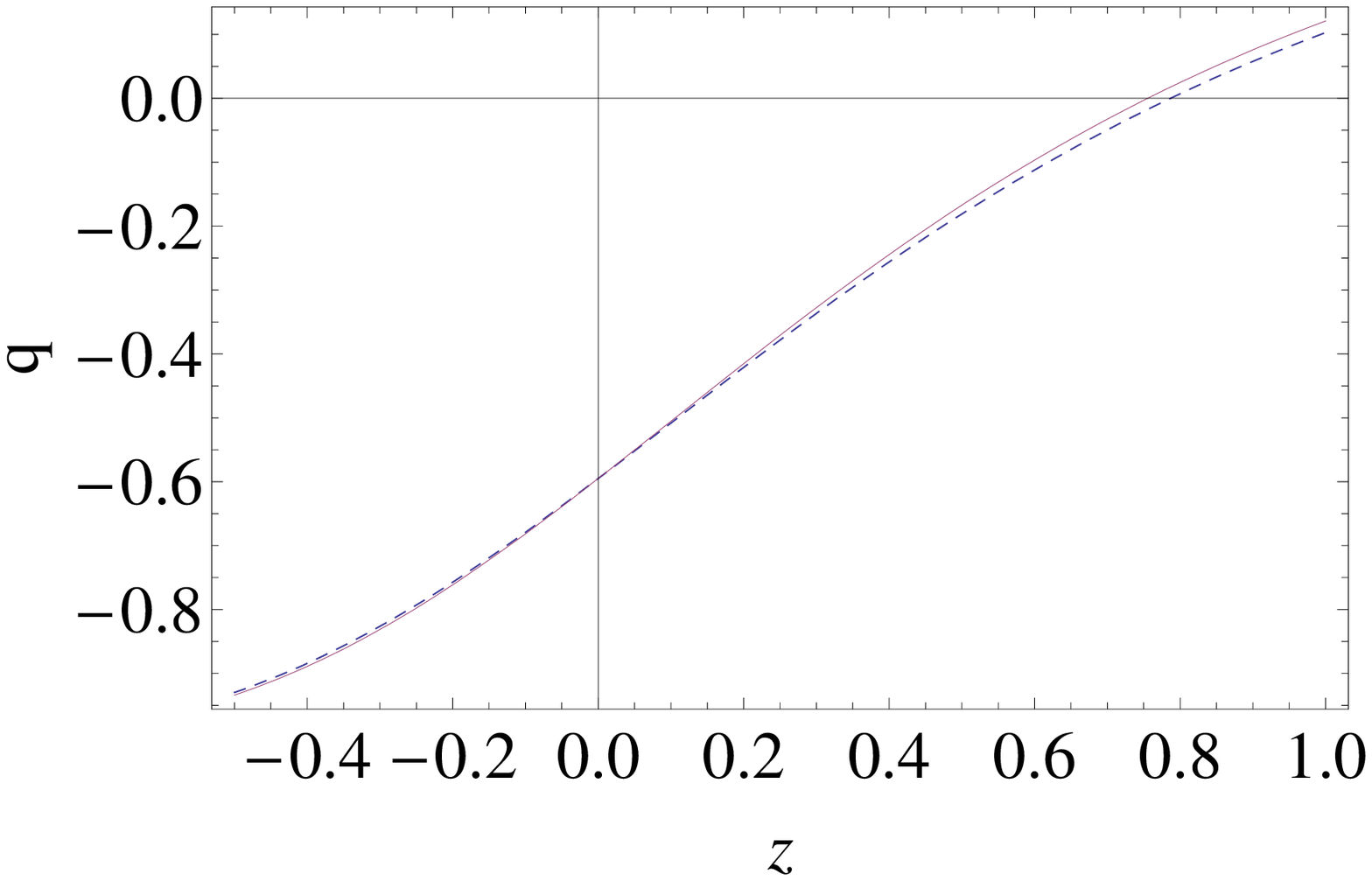}
\includegraphics[height=1.5in]{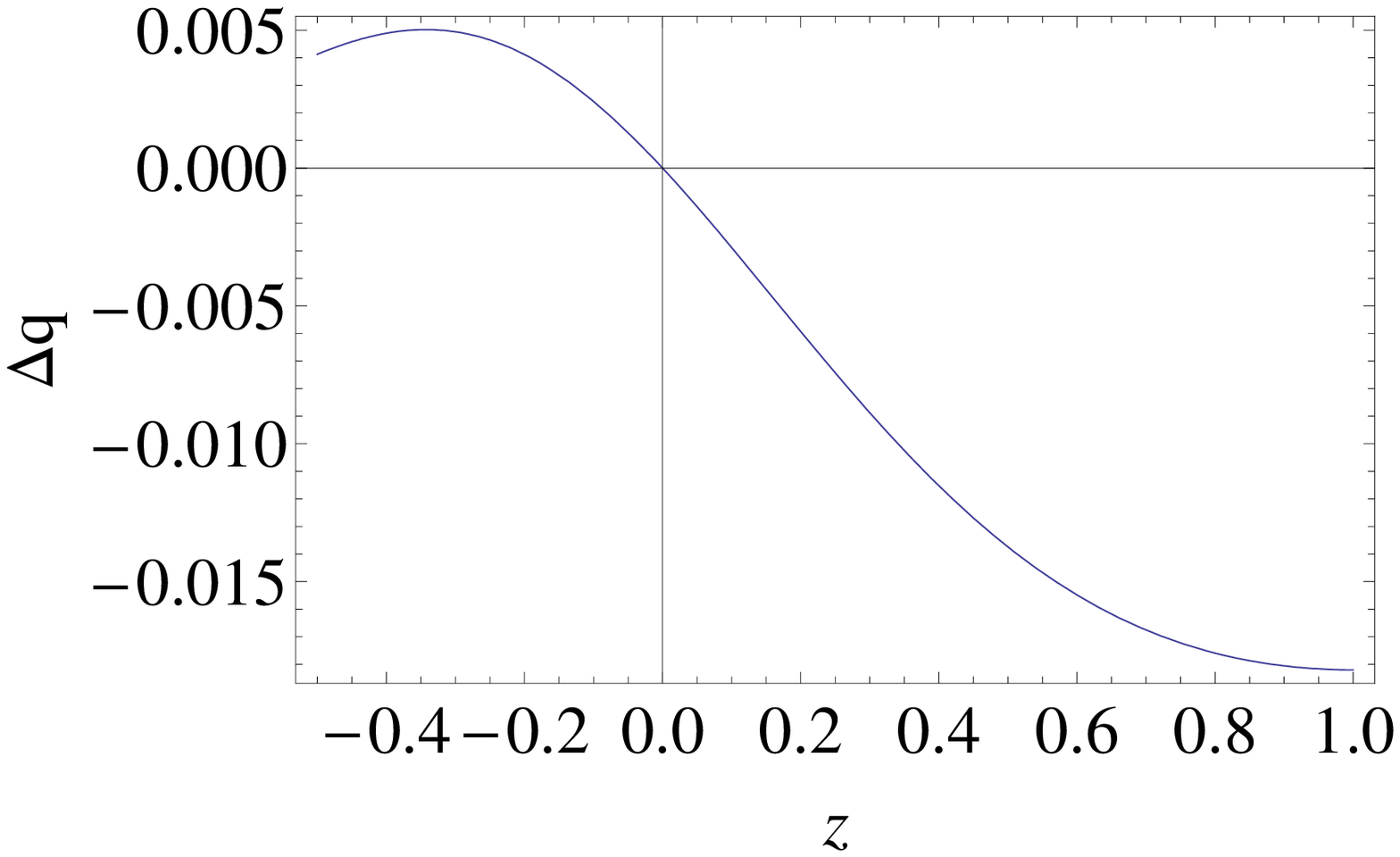}
\caption{We illustrate the deceleration parameter $q$ 
(dashed line) of Eq.~(\ref{ed0}) in contrast 
with its $\Lambda$CDM counterpart (solid line), and their discrepancy (right)
in terms of the redshift $z$.} \label{acceq}
\end{center}
\end{figure}

In this scenario, the time evolutions of gravitational constant 
$G/G_0$ and 
cosmological constant $\Lambda/\Lambda_0$
depend on one parameter $\delta_{_G}$ in Eq.~(\ref{gair0}). 
In order to gain 
some physical insight into these time evolutions, 
and their impacts on the 
cosmological parameters, we chose as an example $\omega_{_M}\approx 0$,
$\delta_{_G}\approx 0.06$ and $k=0$ for illustrations.
In Fig.~\ref{rgl}, $G/G_0$, $\Lambda/\Lambda_0$ and 
$(\Omega_{_\Lambda}/\Omega^0_{_\Lambda})$ 
are plotted in terms of the red 
shift $z\in [-0.5,1]$. It is shown that $G/G_0$ and $\Lambda/\Lambda_0$ (or 
$\Omega_{_\Lambda}/\Omega^0_{_\Lambda})$ are slightly increasing  
and decreasing as the redshift $z$ decreasing, and they are correlated. 
In Fig.~\ref{etimeh}, we plot the Hubble 
function of Eq.~(\ref{re3}) to compare and contrast 
with its $\Lambda$CDM counterpart, and show their discrepancy 
increasing with the redshift $z$, i.e., [$H^2(z),\Delta H^2(z)$].
In Fig.~\ref{acceq}, we plot the deceleration parameter (\ref{ed0}) 
in contrast with its $\Lambda$CDM counterpart, 
and show their discrepancy 
increasing with the redshift $z$, i.e., [$q(z),\Delta q(z)$].  
The same calculations can be done for 
the luminosity distance [$d_H(z),\Delta d_H(z)$].

To end this section, it is interesting to see that in the 
remote future 
$a/a_0\gg 1,\, z\rightarrow -1+0^+$, Eqs.~(\ref{gair}) 
and (\ref{lair}) show that $G/G_0\rightarrow G_{\rm uv}/G_0$ and 
$\Lambda/\Lambda_0\rightarrow \Lambda_c/\Lambda_0 
= (1-\delta_{_\Lambda}/\kappa)\approx 1$ towards the 
UV-stable fixed point 
$(g\rightarrow g_{\rm uv},\tilde\ell\rightarrow \ell_{\rm pl})$. 
Eq.~(\ref{e3}) implies that the universe possibly
undergo a very slow  
``inflation''
\begin{eqnarray}
a\simeq a_0\exp\, (H^2_0\Omega^0_{_\Lambda})^{1/2} 
t\approx a_0\exp\, (t/10^{18}{\rm s}) \,,
\label{sinfl}
\end{eqnarray} 
approaching to  $a_c/a_0=\infty$ in a constant acceleration 
$q\rightarrow -1 + 0^+$. 

\comment{
\section
{\bf A resolution to the coincidence problem}\label{coin}
coincidence and flatness problems. Recall that the cosmological and gravitational 
constants are exactly independent of the time, it is difficult to answer the 
question of cosmological coincidence 
$\Omega^{^0}_{_\Lambda}\sim \Omega^{^0}_{_M}$ in the present time. However, the 
cosmological and gravitational constants evolve in the time, the cosmological 
coincidence $\Omega^{^0}_{_\Lambda}\sim \Omega^{^0}_{_M}$ in the present time is 
just a neutral solution to the Einstein equation with the relationship of the 
cosmological and gravitational constants varying with the time.
}

\section
{\bf The domain of UV-unstable fixed point 
for the universe inflation}\label{inflation}

In the last section, we give a preliminary discussion on 
the possibility whether 
the inflation can occur in the domain
of UV-unstable fixed point $g_{\rm ir}\gtrsim 0$ 
of Eq.~(\ref{gir}). As discussed in Sec.~\ref{irpoint},
this $g_{\rm ir}$-domain is a crossover domain rather than 
the scaling-invariant one, like $g_{\rm uv}$-domain.
In order to gain some physical insight into this domain, 
we assume that $\tilde\xi > \tilde\ell$ at least for some 
part of the $g_{\rm ir}$-domain, and expand 
the coupling $g(\tilde\ell)$ and $\beta$-function (\ref{beta})
as a series 
\begin{eqnarray}
g(\tilde\ell)&=& g_{\rm ir} 
+\tilde c_0(\tilde\ell/\tilde\xi)^{1/\tilde\nu}
+ {\mathcal O}[(\tilde\ell/\tilde\xi)^{2/\tilde\nu}],~~~~(\tilde\xi > \tilde\ell),
\label{gexpf}\\
\beta(g) &=& \beta(g_{\rm ir}) 
+ \beta'(g_{\rm ir}) (g-g_{\rm ir}) + {\mathcal O}[(g-g_{\rm ir})^2]>0,
\label{betaexpf}
\end{eqnarray}
where the coefficient $\tilde c_0>0$, $\beta(g_{\rm ir})=0$ 
and $\beta'(g_{\rm ir})=1/\tilde\nu >0$.  
In the $g_{\rm ir}$-domain,
the behavior of the $\beta$-function is
\begin{eqnarray}
\beta'(g_{\rm ir})(g-g_{\rm ir})>0,~~~~~~{\rm for}~~~~ g>g_{\rm ir}\gtrsim 0
\label{fbetaf}
\end{eqnarray}
indicating the fixed point $g_{\rm ir}$ is UV-unstable, 
the coupling $g(t)$ moves away from $g_{\rm ir}$,
as the UV-cutoff $\tilde\ell(t)$ 
decreases. 

In this $g_{\rm ir}$-domain, the scaling law for $\tilde\xi\gg \tilde\ell$ 
is not completely valid and the action is more
complicate than the Einstein action (\ref{ec0}).
Nevertheless, as a preliminary study and for the reasons discussed in Sec.~\ref{irpoint}, 
we approximately use Eq.~(\ref{rge})
for $m=\tilde\xi^{-1}$ and obtain
\begin{eqnarray}
\tilde \xi (t)&\approx & \tilde\ell(t)\exp +\int_{g_{\rm ir}}^{g(t)}\frac{dg'}{\beta(g')},
\quad {\rm for}\quad g(t)>g_{\rm ir},\nonumber\\
&=&\tilde\ell(t)\Big[\tilde c_0[g(t)-g_{\rm ir}]\Big]^{\tilde\nu/2},
\label{irf}
\end{eqnarray}
and the dimensionless scaling factor
\begin{eqnarray}
\tilde a(t)\propto \tilde\xi(t)/\tilde\ell(t)
=\Big[\tilde c_0[g(t)-g_{\rm ir}]\Big]^{\tilde\nu/2},
\quad {\rm for}\quad g(t)>g_{\rm ir}.
\label{iraf}
\end{eqnarray}
Suppose that the initial scaling factor $\tilde a_{_0}$ and 
gravitational coupling 
$\tilde g_{_0}=g(\tilde a_{_0})\gtrsim g_{\rm ir}$, where  
a smaller subscript or 
superscript ``${_0}$'' is used to indicate quantities' 
values at the initial time $t_{_0}$, 
differently from the normal subscript or superscript ``0'' indicating 
quantities' values at the present time $t_0$. 
The initial energy densities are
\begin{eqnarray}
\tilde \rho^{^0}_{_{\tilde\Lambda}}=\tilde \Lambda_{_0}/(8\pi \tilde G_{_0}),
~~~\tilde \rho^{^0}_{_M}\approx 1/\tilde\xi^4_{_0},
~~~\tilde \rho^{^0}_k=-k/(8\pi \tilde G_{_0} \tilde\xi^2_{_0}), 
\label{dinf}
\end{eqnarray} 
where $\tilde \Lambda_{_0}\propto \tilde\xi^{-2}_{_0}$ 
of Eq.~(\ref{cmc}) and $\tilde G_{_0}=\tilde g_{_0} G_0
= \tilde g_{_0} \ell_{\rm pl}^2 \gtrsim 0$. 
The space-time entropy 
$\tilde S^{^0}_{_\Lambda}\propto\tilde a_{_0}$ and energy 
$\tilde E^{^0}_{_\Lambda}\propto \tilde a_{_0}\tilde\ell^{-1}(t_{_0})$ 
are given by Eqs.~(\ref{ent}) and (\ref{eng}).
Analogously to Eq.~(\ref{omnow}), we define 
\begin{eqnarray}
 \tilde \rho^{^0}_c=3\tilde H^2_{_0}/(8\pi \tilde G_{_0}),
\quad \tilde \Omega^{^0}_{_{M,\tilde\Lambda,k}}
=\tilde\rho^{^0}_{_{M,\tilde\Lambda,k}}/\tilde \rho_c^{^0},
\label{crf}
\end{eqnarray} 
where $\tilde \rho^{^0}_c$ is the critical density, 
the Hubble rate $\tilde H=\dot{\tilde a}(t)/\tilde a(t)$ 
and $\tilde H_{_0}=\dot{\tilde a}(t_{_0})/\tilde a(t_{_0})$. 

In the $g_{\rm ir}$-domain,
we consider the universe evolution in  
$g \gtrsim \tilde g_{_0} \gtrsim g_{\rm ir}$ 
and $\tilde a> \tilde a_{_0} > \tilde a_{\rm ir}$
\comment{
\begin{description}
\item (i) $g \gtrsim \tilde g_{_0} \gtrsim g_{\rm ir}$ 
and $\tilde a> \tilde a_{_0} > \tilde a_{\rm ir}$;
\item (ii) $\tilde g_{_0} \gtrsim g \gtrsim g_{\rm ir}$ 
and $\tilde a_{_0} > a > \tilde a_{\rm ir}$,
\end{description}
}
where $\tilde a_{\rm ir}$ is the scaling factor 
for $g_{\rm ir}=g(\tilde a_{\rm ir})$.   
For two different values 
$\tilde a$ and $\tilde a_{_0}$ of the scaling factor, Eq.~(\ref{iraf}) yields 
\begin{eqnarray}
\tilde a^{2} \propto (\tilde c_0)^{\tilde\nu}(g-g_{\rm ir})^{\tilde\nu};\quad
\tilde a^{2}_{_0} \propto   (\tilde c_0)^{\tilde\nu}(\tilde g_{_0}-g_{\rm ir})^{\tilde\nu},
\label{ir1f}
\end{eqnarray}
and the ratio of these two equations leads to the running gravitational coupling 
as a function of the scaling factor
\begin{eqnarray}
\left(\frac{g}{\tilde g_{_0}}\right) 
&=&\left(\frac{g_{\rm ir}}{\tilde g_{_0}}\right) 
+ \left(1-\frac{g_{\rm ir}}{\tilde g_{_0}}\right)
\left(\frac{ \tilde a}{\tilde a_{_0}}\right)^{2/\tilde\nu}\label{gairf}\\
&\approx & \left(\frac{\tilde a}{\tilde a_{_0}}\right)^{2/\tilde\nu},
~~~~~~~~~~~~~~~g_{\rm ir}\gtrsim 0 \label{gairf0}
\end{eqnarray}
where $[G(t)/\tilde G_{_0}]=[g(t)/\tilde g_{_0}]$.

\comment{
Similarly to Eq.~(\ref{betaexp}), the $\beta$-function (\ref{beta}) 
around the fixed point $g_{\rm ir}$ be given by a series
\begin{eqnarray}
\beta(g) &=& \beta(g_{\rm ir}) + \beta'(g_{\rm ir}) g + {\mathcal O}(g^2),
\label{betaexp1}
\end{eqnarray}
where $\beta(g_{\rm ir})=0$ and $\beta'(g_{\rm ir})=1/\tilde\nu>0$.
We integrate Eq.~(\ref{betaexp1}) and obtain 
\begin{eqnarray}
[G(t)/\tilde G_{_0}]&=&[g(t)/\tilde g_{_0}]=[\tilde\ell(t)/\tilde\xi_{_0}]^{-1/{\tilde\nu}}, 
\label{gauv}
\end{eqnarray}
where $\tilde\nu\approx 1/2$ is the perturbation result \cite{w6}. 
In Eq.~(\ref{gauv}), 
the integration constants 
\begin{eqnarray}
\tilde g_{_0}=\tilde G_{_0}/G_0\gtrsim g_{\rm ir}\approx 0,
\quad \tilde\xi_{_0}>\ell_{\rm pl}, 
\label{ginf}
\end{eqnarray} 
are the characteristic coupling and length 
in the domain of the UV-unstable fixed point 
$g_{\rm ir}$. 
}
 
In order to gain a physical insight into the possibility whether or not 
the inflation can take place  in the $g_{\rm ir}$-domain,
we assume the decoupling of the cosmological term  
and matter fields, and approximately adopt Eq.~(\ref{geqi0}) or (\ref{geqi2}).    
We write Eq.~(\ref{geqi2}) as
\begin{eqnarray}
\frac{d\tilde\Lambda}{dt}
&=&-8\pi\rho_{_M}\frac{dG}{dt}, ~~ {\rm or}~~
\frac{d(G\tilde\rho_{_\Lambda})}{d\tilde x}=-\tilde\rho_{_M}\frac{dG}{d\tilde x},
\label{geqi2f}
\end{eqnarray}
where $\tilde x(t)=\tilde a(t)/ \tilde a_{_0}(t_{_0})$ and 
$\tilde x_{_0}(t_{_0})=1$.
Then integrating Eq.~(\ref{gairf}) over 
``time''-variable $\tilde x\equiv \tilde a(t)/\tilde a_{_0}$ in the region 
$\tilde x > 1$ with the boundary 
condition $\tilde x(t_{_0})=1$, we obtain
\begin{eqnarray}
(\tilde\Lambda/\tilde\Lambda_{_0})
&=&(G\tilde\Omega_{_{\tilde\Lambda}})/(\tilde G_{_0}\tilde\Omega^{^0}_{_{\tilde\Lambda}})\nonumber\\
&=& 1-(\tilde \delta_{_\Lambda} /\tilde \kappa)
\left[1 - \left(\frac{\tilde a_{_0}(t_{_0})}{
\tilde a(t)}\right)^{\tilde \kappa}\right],
\label{lauv0}
\end{eqnarray}
where the parameters are 
\begin{eqnarray}
\tilde \kappa &=&3(1+\omega_{_M})-1/\tilde\nu >0,\quad
\tilde \delta_{_\Lambda} 
= (1/\tilde\nu) (\tilde \Omega^{^0}_{_M}/\tilde \Omega^{^0}_{_{\tilde\Lambda}})
>0,
\label{dexp}
\end{eqnarray}
and $\omega_{_M}\approx 1/3$ for ultra relativistic particles.  

In the initial condition, it is compellingly reasonable to assume that the
matter-field energy-density $\tilde \rho^{^0}_{_{M}}$
is much smaller than the space-time
energy-density $\tilde \rho^{^0}_{_{\tilde\Lambda}}$, see 
Eq.~(\ref{dinf}), so that
$(\tilde \Omega^{^0}_{_M}/\tilde \Omega^{^0}_{_{\tilde\Lambda}})
=(\tilde \rho^{^0}_{_M}/\tilde \rho^{^0}_{_{\tilde\Lambda}})\ll 1$. 
Suppose that $\tilde\nu\gg 1$ 
and $\beta'(g_{\rm ir})=1/\tilde\nu \ll 1$, namely, 
the $\beta$-function is very flat  in the $g_{\rm ir}$-domain,
see also the discussion presented in the last paragraph of Sec.~\ref{irpoint}. 
As a result, $\tilde \delta_{_\Lambda}/\kappa \ll 1$ in 
Eq.~(\ref{lauv0}), $\tilde\Lambda\approx \tilde\Lambda_{_0}$ 
slowly varies. Eq.~(\ref{e3}) becomes
\begin{eqnarray}
\tilde H^2
=\tilde H_{_0}^2(G/\tilde G_{_0})
(\tilde \Omega_{_M} + \tilde \Omega_{_\Lambda}+\tilde \Omega_{_k})
\approx (\tilde \Lambda-k\tilde a^{-2})/3,\label{ue3}
\end{eqnarray}
which shows an inflationary de Sitter solution, 
$\tilde \Lambda_{_0}\approx 3\tilde H_{_0}^2$, 
\begin{eqnarray}
\tilde a(t) \approx \tilde a(t_{_0}) \exp\, (\tilde\Lambda_{_0}/3)^{1/2} (t-t_{_0})\,. 
\label{dest}
\end{eqnarray}
Equation (\ref{gairf}) gives the coupling
\begin{eqnarray}
\tilde g(t) \approx \tilde g(t_{_0}) 
\exp\, (2/\tilde\nu)(\tilde\Lambda_{_0}/3)^{1/2} (t-t_{_0})\,, 
\label{destg}
\end{eqnarray}
which demands $\beta'(g_{\rm ir})=1/\tilde\nu \ll 1$ to be consistent 
with small variation of the coupling $g$ (\ref{gairf}). The space-time entropy  
$\tilde S_{_\Lambda}\propto 
\tilde a(t)$ and energy
$\tilde E_{_\Lambda}\propto \tilde a(t)\ell^{-1}(t)$
of Eqs.~(\ref{ent}) and (\ref{eng}) increase. 

The proliferation and increase of space-time entropy 
$\tilde S_{_\Lambda}$ drive the universe inflation (\ref{dest}). 
Moreover, the entropy of particles and antiparticles increases by converting 
the space-time energy-density 
$\tilde \rho_{_{\tilde\Lambda}}$ to the matter-field energy-density 
$\tilde \rho_{_{M}}$.  In the initial inflation epoch 
the space-time energy-density $\tilde \rho_{_{\tilde\Lambda}}$ 
is much more larger and energetic 
for the production of particles and antiparticles.

Similarly to the relation (\ref{omnow}) and 
$\rho^{0}_{_{\rm total}}=\rho^{0}_c$ in the present time $t_0$,
Eqs.~(\ref{crf}) and (\ref{ue3}) give 
$(\tilde \Omega^{^0}_{_M}+\tilde \Omega^{^0}_{_\Lambda}
+\tilde \Omega^{^0}_{_k})=1$, indicating the total 
energy-density $\tilde\rho^{^0}_{_{\rm total}}=\tilde\rho^{^0}_c$ 
in the initial time $t_{_0}$, although 
the initial values $\tilde a_{_0}$, $\tilde\xi_{_0}$, 
$\tilde g_{_0}$ and densities (\ref{dinf}) are unknown.
Needless to say, in order to understand the quantitative 
properties of entire inflationary 
process, for example, whether the inflation ends 
with the $e$-folding factor 
$\ell n[\tilde a(t)/\tilde a_{_0}(t_{_0})]\approx 60$, 
we have to study the rate of
the energy-conversion and particle-creation 
form the the space-time energy-density 
$\tilde \rho_{_{\tilde\Lambda}}$ to the matter-field energy-density 
$\tilde \rho_{_{M}}$, and adopt the generalized 
Bianchi identity (\ref{geqi00}) or (\ref{cgeqi2}). 
The initial values $\tilde a_{_0}$, $\tilde\xi_{_0}$, 
$\tilde g_{_0}$ and densities (\ref{dinf}), as well as the value $1/\tilde\nu$ 
could be determined by either theories or observations.

In addition to the problem previously discussed, there are 
many other open questions. We mention a few more examples. Actually, 
the correlation length $\xi(t)$ is much larger than the UV-cutoff
$\tilde\ell(t)$ in the entire universe evolution from the $g_{\rm ir}$-domain 
to the $g_{\rm uv}$-domain. After the crossover domain, we 
speculate only one 
scaling law $a(t)\propto[\xi(t)/\tilde\ell(t)]=\xi[g(t)]$, 
provided the full $\beta$-function $\beta(g)$ is known 
in the entire range of the coupling $g(t)\in (g_{\rm ir}, g_{\rm uv}]$. 
Moreover, induced by very massive 
torsion fields, the 
four-fermion interactions depend on 
the gravitational gauge coupling $g$
in the Einstein-Cartan theory \cite{xue4f2015}. 
How these self-interactions of matter 
fields affect on the universe evolution. 
There unsolved issues deserve studies in future.

\noindent
{\bf Acknowledgment.}  
\hskip0.1cm 
Author thanks 
Prof.~Remo Ruffini for discussions on the Einstein 
theory and cosmology. 

\end{document}